\documentclass[prd,preprint,showpacs,preprintnumbers,nofootinbib]{revtex4}

\usepackage{amsmath}
\usepackage{amssymb}
\usepackage{graphicx}
\usepackage[tight]{subfigure}
\usepackage[usenames]{color}
\usepackage{txfonts}

\newcommand{\Tr}{{\rm Tr}} 
\newcommand{\sla}[1]{{#1\hskip-0.45em/}}

\newcommand{\SO}{{\rm SO}}
\newcommand{\SU}{{\rm SU}}
\newcommand{\U}{{\rm U}}

\begin{document}


\preprint{RIKEN-TH-70}
\preprint{hep-th/0604211}


\title{Improved Perturbation Method and its Application to the IIB Matrix Model}

\author{T.~Aoyama}
\affiliation{Theoretical Physics Laboratory, RIKEN, Wako, 351-0198, Japan }

\author{Y.~Shibusa}
\affiliation{Theoretical Physics Laboratory, RIKEN, Wako, 351-0198, Japan }

\begin{abstract}
We present a new scheme for extracting approximate values 
in ``the improved perturbation method'', which is a sort of 
resummation technique capable of evaluating a series 
outside the radius of convergence. 
We employ the distribution profile of the series that is 
weighted by $n$th-order derivatives with respect to the 
artificially introduced parameters. 
By those weightings the distribution becomes more sensitive 
to the ``plateau'' structure 
in which the consistency condition of the method is satisfied. 
The scheme works effectively even in such cases that 
the system involves many parameters. 
We also propose that this scheme has to be applied to each 
observable separately and be analyzed comprehensively. 

We apply this scheme to the analysis of the IIB matrix model 
by the improved perturbation method 
obtained up to eighth order of perturbation in the former works. 
We consider here the possibility of spontaneous breakdown 
of Lorentz symmetry, and evaluate the free energy and the 
anisotropy of space-time extent. 
In the present analysis, we find an $\SO(10)$-symmetric vacuum 
besides the $\SO(4)$- and $\SO(7)$-symmetric vacua that have 
been observed. 
It is also found that there are two distinct $\SO(4)$-symmetric 
vacua that have almost the same value of free energy but 
the extent of space-time is different. 
From the approximate values of free energy, we conclude that 
the $\SO(4)$-symmetric vacua are most preferred among those 
three types of vacua. 
\end{abstract}



\maketitle


\newpage

\section{Introduction\label{sec:intro}}

String theory is the unique theory that contains massless spin-two 
particles, i.e. gravitons
\cite{Yoneya:1974jg}, 
and thus it is considered to provide an unified microscopic 
description of the universe including gravitational interactions. 
For this reason the string theory has been subjected to 
intensive studies. 
However, it is recognized that 
the perturbative string theory fails to single out 
our universe as the unique vacuum of the theory
\cite{Kawai:1986ah}. 
Therefore we are forced to pursue non-perturbative formulations. 
The IIB matrix model (also called the IKKT matrix model) 
is proposed as a constructive formulation of the superstring theory 
\cite{Ishibashi:1996xs,Aoki:1998bq}.

A significant feature of the IIB matrix model is that the space-time 
itself is expressed by the eigenvalue distribution of 10 bosonic 
matrices, and thus it is treated as a dynamical variable of the model. 
The origin of our four-dimensional space-time can be argued 
in the context of the IIB matrix model as a spontaneous breakdown 
of Lorentz symmetry. In this regard, we have to understand 
non-perturbative properties of the model. 

The mechanism of the spontaneous breakdown of Lorentz symmetry 
in reduced matrix models has been examined in various approaches. 
It has been recognized from those works that 
the fermionic part of the action plays a crucial role 
\cite{%
Hotta:1998en,Ambjorn:2000dx,Ambjorn:2001xs,%
Nishimura:2000wf,Nishimura:2000ds,%
Nishimura:2001sq,%
Anagnostopoulos:2001yb}.

Unveiling dynamical aspects of the model is, in general, 
quite a difficult problem. 
The Monte-Carlo method is a powerful tool for exploring such 
non-perturbative properties of a model. 
However, it is not applicable (or at least quite difficult to apply) 
to the IIB matrix model due to complex phase of the action derived 
from the fermionic part\footnote{%
A novel technique called the factorization method is proposed 
to resolve the complex action problem in the Monte Carlo simulations 
\cite{Anagnostopoulos:2001yb,Ambjorn:2002pz}.
}. 
The improved perturbation method (also called the Gaussian expansion 
method) is an alternative approach. 
It is considered as a sort of variational method 
\cite{%
Yukalov,Caswell:1979qh,Halliday:1979xh,%
Stevenson:1981vj,Killingbeck,Dhar:1982sh}.
It has been successfully applied to various models 
\cite{%
Guida:1994zv,Guida:1995px,%
Kawamoto:2003kn,Aoyama:2005nd},
and applications to matrix models were done in 
Refs.~\cite{Kabat:1999hp,Oda:2000im,Sugino:2001fn}.

The application to the IIB matrix model was first achieved in 
Ref.~\cite{Nishimura:2001sx}
in which various patterns of symmetry breaking that preserve
$\SO(d)$ subgroup of the original ten-dimensional rotational symmetry 
(ansatz) were examined, 
and as a conclusion, four-dimensional universe is the most preferred 
among them based on the comparison of free energy. 
The Gaussian expansion method was reformulated as 
an improvement of perturbative series expansion in 
Ref.~\cite{Kawai:2002jk}. 
The improved Taylor expansion (ITE), as it is referred, 
opened a way toward more general applications 
that incorporate quadratic and other types of interactions. 
The ITE prescription was employed for the IIB matrix model 
up to fifth order of perturbation in Ref.~\cite{Kawai:2002jk}. 
It has been proceeded to even higher orders and extended ansatz 
\cite{Kawai:2002ub,Aoyama:2006rk,Aoyama:2006di}.
The mechanism of the spontaneous symmetry breakdown 
is further examined in a simplified model 
via the Gaussian expansion method in 
Ref.~\cite{Nishimura:2004ts}.

In the present paper, we investigate the non-perturbative 
solutions of the IIB matrix model by this technique 
(which we call ``the improved perturbation method'' here). 
We focus on the possibility 
that the original $\SO(10)$ symmetry of the IIB matrix model 
may be spontaneously broken to result in our universe 
which spreads in four directions and has $\SO(4)$ rotational 
symmetry\footnote{%
We perform the Wick rotation to the IIB matrix model and discuss in 
Euclidean space-time.}. 
In such cases it is important to see that the method is applicable 
to the models that exhibit phase transitions. 
In the former work \cite{Aoyama:2005nd}, 
we applied this method to the Ising model and found that 
the improved perturbation method extracts the information of 
the ordered phase from an expansion about the vacuum in 
the disordered phase. 
It is also observed that an unstable vacuum is identified 
even though it has larger value of free energy than that of the 
stable vacuum. 
We expect that the improved perturbation method reveals different 
patterns of symmetry breaking that may be developed as unstable 
or metastable vacua. 

The improved perturbation method is considered as a sort of 
resummation of perturbative series by introducing the artificial 
parameters into the model. 
The approximate value of the series is obtained by evaluating it 
in the region of the artificial parameter space that realizes 
the principle of minimal sensitivity \cite{Stevenson:1981vj}. 
We call such a region as ``plateau'', in which the dependence 
on those parameters would vanish effectively 
and the exact value would be reproduced. 
It works as a consistency condition for the parameters. 

The concept of plateau is rather obscure, and there is not yet 
any proper treatment of mathematical rigor. 
We need a practical scheme for extracting the approximate value 
of the series. 
In addition, it should be free from the ambiguity due to 
the subjectivity of recognition as much as possible. 
In the former works, mainly two approaches have been propounded. 
One is to take the values at the extrema of the improved series 
with respect to the artificial parameters as approximate values 
\cite{Nishimura:2001sx}. 
The other is to take the values which corresponds to the mode of 
distribution of the evaluated values around the accumulation of 
extrema. The latter is called the histogram technique 
\cite{Nishimura:2002va,Nishimura:2003gz}. 

In this article we propose a new prescription which enables us to extract 
a good approximation from the improved series. 
It is achieved by incorporating the appropriate weightings in the histograms 
that are given by the inverse of derivatives. 
This method will be applicable to various models which have many
parameters. 

It has also been customary to evaluate the various observables 
based on the information of plateau of the free energy. 
On extracting the approximate value of any observable, 
the information of minimal sensitivity of only improved free energy is 
taken into account. 
To be more precise, one takes the value of improved perturbative 
series of an observable at extrema of improved free energy with respect 
to the artificial parameters as approximate values of that observable. 
We insist that the improved perturbation method has to be 
applied independently for each observable of interest. 

We apply the new scheme to the analysis of the IIB matrix model 
by means of the improved perturbation method. 
It is based on the improved series that is obtained 
in former works up to eighth order of perturbation 
for $\SO(d)$-preserving configurations as ansatz ($d=4$ and $7$)
\cite{Aoyama:2006rk}. 
 
This paper is organized as follows. 
In Section~\ref{sec:optimized}, 
we provide a description of the resummation technique called 
the improved perturbation method, including a review of the method. 
In Section~\ref{sec:plateau}, 
we propose a new prescription which extracts the approximate 
value from improved perturbative series. 
In Section~\ref{sec:ikkt}, 
we apply the improved perturbation to the IIB matrix model 
and survey the non-perturbative solutions with various patterns of 
symmetry breaking. 
Section~\ref{sec:conclusion} is devoted to the conclusion. 
In Appendix~\ref{sec:detail}, we present a detail of analysis 
performed in Section~\ref{sec:ikkt}. 


\section{Improved Perturbation\label{sec:optimized}}

It is generally believed that the perturbative expansion 
(also including $\frac{1}{N}$-expansion and $\varepsilon$-expansion) 
is asymptotic, and diverges beyond some finite orders.
Nevertheless, what we can evaluate for most of the theories are 
only those series expansions, from which we have to draw physical 
information of the theory.
Therefore, we need a method to estimate the {\em exact} value from 
the series with the parameter (coupling constants, and so on) 
outside the original convergence radius.
For this purpose we introduce here a method which has been successfully 
applied to the IIB matrix model and other general models.

\subsection{Prescription}

We implicitly use $\lambda$ as coupling constant and $m$ as 
parameters of the model collectively such as masses.
Let us assume that the observables of a theory would be exactly described by a 
function $F(\lambda, m)$.
Perturbation theory provides an expansion of $F$ as a power 
series of $\lambda$ about $\lambda = 0$, 
with $n$th coefficient denoted as $f_n(m)$:
\begin{equation}
	F(\lambda, m) 
	= \sum_{n=0}^{\infty}\,\lambda^n\,f_{n}(m) \,.
\end{equation}
In the actual cases, we only have finite portion of the series 
up to order $N$, 
\begin{equation}
	F^N(\lambda, m) 
	= \sum_{n=0}^{N}\,\lambda^n\,f_{n}(m) .
\end{equation}
The question is whether we can presume the {\it exact} value 
$F(\lambda, m)$ at the given parameters $\lambda$ and $m$ 
from the series $F^N$ above. 
In many cases the convergence radius for $\lambda$ is zero, so 
we can not expect that $F^N$ gives a reliable approximation.

Now we consider a modification of the series along the following 
prescriptions. First we perform a shift of parameters:
\begin{equation}
	\begin{aligned}
	\lambda &\longrightarrow g^p\,\lambda \,, \\
	m       &\longrightarrow m_0 + g^q (m - m_0) ,
	\end{aligned}
\label{eq:improve_parameter}
\end{equation}
where we have introduced $g$ as a formal expansion parameter, 
and $m_0$ as a set of artificial parameters. 
$p$ and $q$ are taken arbitrarily. 
We deform the series by the substitution (\ref{eq:improve_parameter}), 
and then we reorganize the series in terms of $g$, 
drop the ${\cal O}(g^{M+1})$ terms, 
and finally set $g$ to 1. 
We obtain the improved perturbative series $\widetilde{F}^N$ as
\begin{equation}
	F^N (\lambda, m) \,
	\longrightarrow \,
	\widetilde{F}^N (\lambda, m; m_0)
	=
	F^N (g^p\lambda, m_0+g^q(m-m_0))\Bigr|_{g^M, g\rightarrow 1}. 
\end{equation}
Here, we adopt a notation $\bigr|_{g^M,\,g\rightarrow 1}$ 
to represent the operation that 
we disregard the ${\cal O}(g^{M+1})$ terms and then put $g$ to 1.

It should be noted that by simply setting $g=1$, the modification 
itself becomes trivial and the series would be independent of the 
parameters $m_0$. 
However, due to dropping the ${\cal O}(g^{M+1})$ terms, 
the deformed series does depend on $m_0$.

To turn the argument around, we adopt here the principle of 
minimal sensitivity that the {\em exact} value $F$ will be 
reproduced when the improved series depends least on the 
artificial parameters $m_0$. 
It provides a sort of consistency condition on $m_0$; by tuning 
the parameters to the solution of the condition, we will have 
a good approximation of $F$.

In the above we have introduced the arbitrary parameters $p,q,M$. 
If the region emerges in the parameter space of $m_0$ 
that realizes minimal sensitivity, 
the approximate value must be independent of the choice of 
parameters $p,q,M$. 
However, with the limited orders of perturbation, 
the signal of minimal sensitivity is often weak in actual cases. 
So we have to find the optimal values of parameters $p,q,M$ in 
order to make the signal clear. 
We have to keep in mind that large values of $p$ and $q$ turn 
to throw away the information of higher order terms of perturbative 
series, and taking larger value of $M$ than that of $N$ is meaningless. 
Therefore, we set the parameters as $p=q=1$ and $M=N$ throughout this paper. 

\subsection{Example}

For example, we apply the improved perturbation method to a simple 
1-dimensional function,
\begin{equation}
	F(\lambda, m) = \frac{1}{1+\lambda m} \,,
\end{equation}
and try to estimate $F(\lambda, m)$ at $\lambda=1$ and $m=3/2$.

\begin{figure}
\begin{center}
\subfigure[][]{%
	\includegraphics[scale=.75]{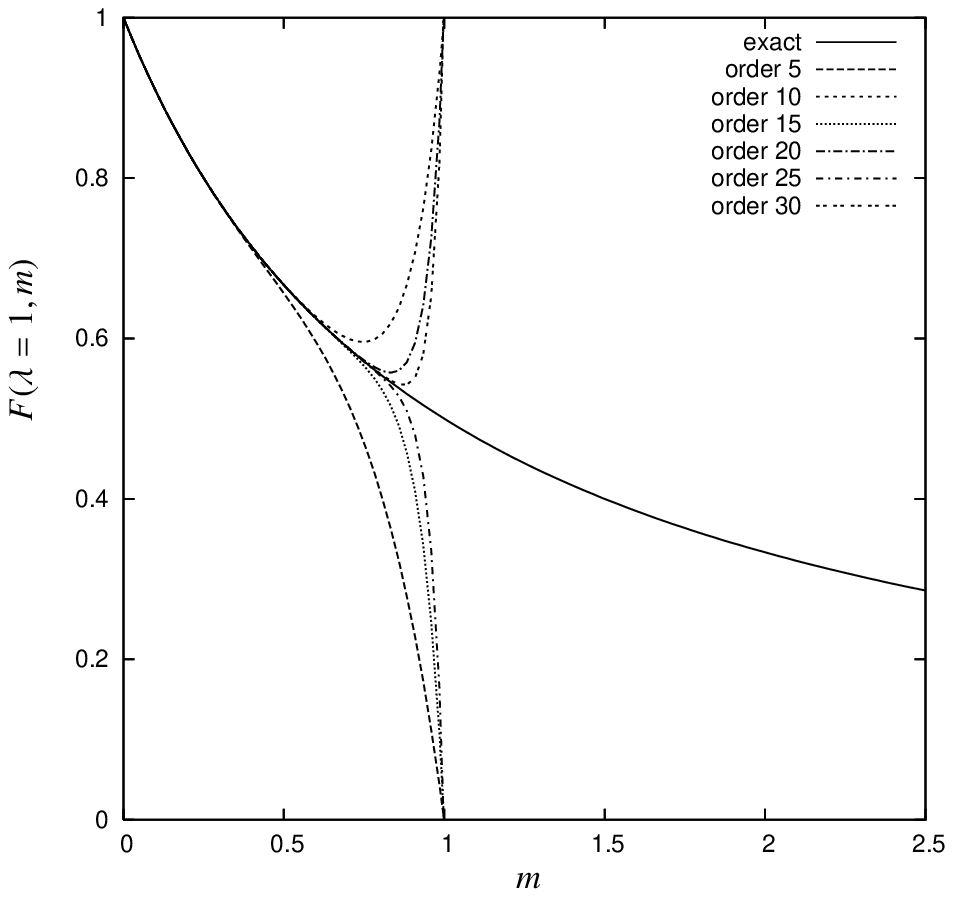}
	\label{fig:fun01-1}
}
\hskip 1em
\subfigure[][]{%
	\includegraphics[scale=.75]{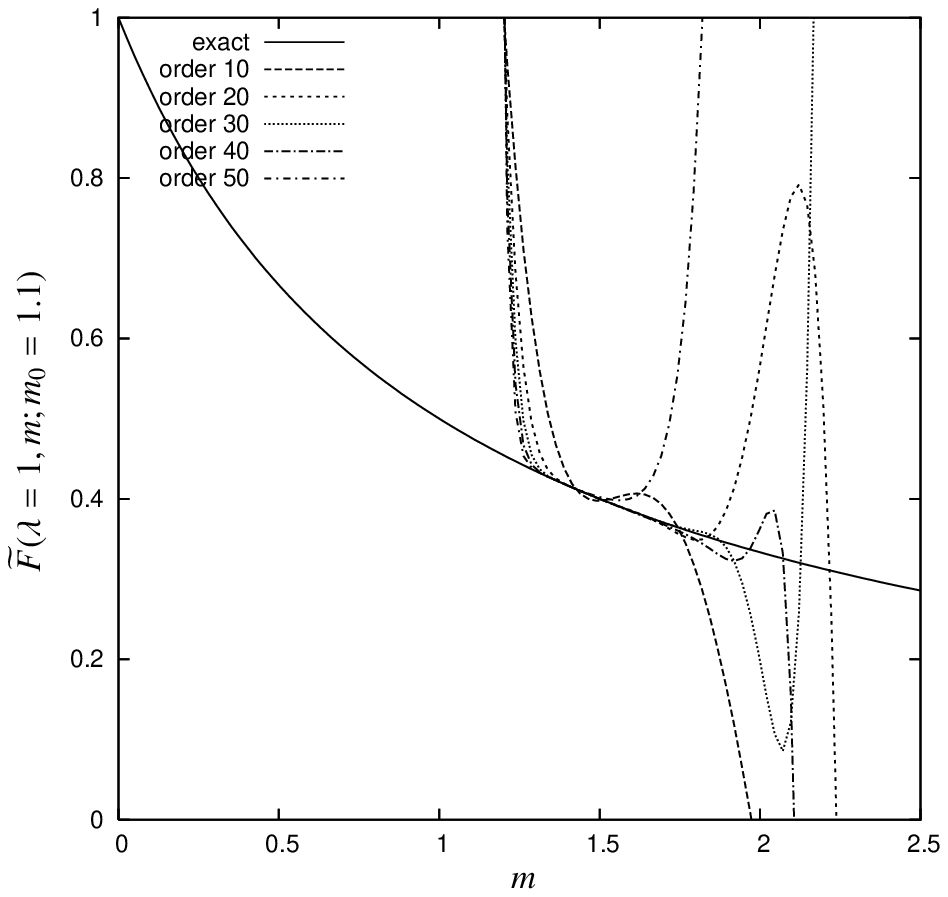}
	\label{fig:fun01-2}
}
\end{center}
\caption{%
\subref{fig:fun01-1} Taylor series of $F(\lambda, m)=\frac{1}{1+\lambda m}$ 
and \subref{fig:fun01-2} improved series at $m_0 = 1.1$.}
\label{fig:fun01}
\end{figure}

The Taylor expansion about $\lambda = 0$ up to order $N$ gives 
the following series:
\begin{equation}
	F^N(\lambda, m) = \sum_{n=0}^{N}\,(-1)^n\,\lambda^n\,m^n \,.
\end{equation}
This series at $N\rightarrow\infty$ has finite convergence radius 
$|\lambda\,m| < 1$. 
The situation is depicted in Fig.~\ref{fig:fun01-1} where $F^N$ of various $N$ 
are shown as a function of $m$ when $\lambda = 1$.

%
%

Now we deform the series along the prescription in the previous 
section, to obtain the improved series $\widetilde{F}^N$:
\begin{equation}
	\widetilde{F}^N (\lambda, m; m_0)
	= 
	\sum_{n=0}^{N}\,(-1)^n\,
	\left(
		\lambda g)^n\,(m_0+g(m-m_0)
	\right)^n\,
	\Bigr|_{g^N,\,g\rightarrow 1} \,.
\end{equation}
Here, $\bigr|_{g^N,\,g\rightarrow 1}$ denotes disregard for 
the ${\cal O}(g^{N+1})$ terms followed by setting $g=1$.

The improved series behaves as shown in Fig.~\ref{fig:fun01-2}, 
when the artificial parameter $m_0$ is taken to be 1.1. 
In this case, the functions $\widetilde{F}^N(\lambda=1,m;m_0=1.1)$ 
become close to the {\em exact} value 
in the region $m \sim 1.5$, and we will have a good approximation of 
the original function $F$ at $m=3/2$. 

To find the optimum of parameter $m_0$, we consider 
$\widetilde{F}^N (\lambda, m; m_0)$ 
as a function of $m_0$ with $m$ and other original parameters 
fixed to the specified values, 
({\it e.g.} $\lambda=1$ and $m=3/2$ in this case).
%
%
\begin{figure}
\includegraphics[scale=.8]{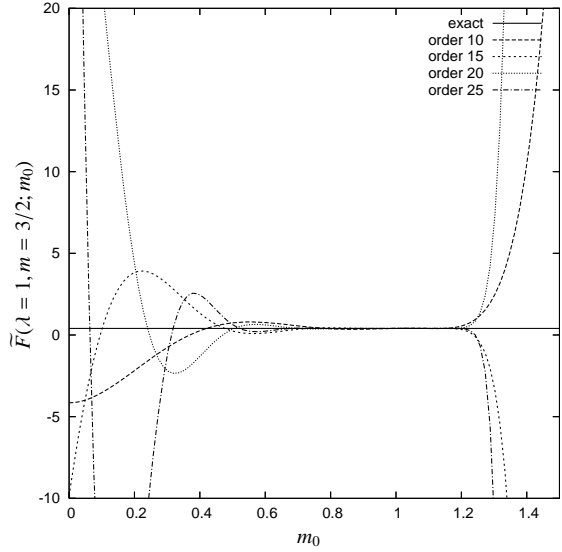}
\caption{Improved series as a function of artificial parameter $m_0$.}
\label{fig:fun01impr}
\end{figure}
It can be seen that in some region of parameter 
$m_0 = 0.7 \sim 1.1$, $\widetilde{F}^N$ stays stable and gives
good approximation to the {\em exact} value, $F=0.4$ (Fig.~\ref{fig:fun01impr}). 
We call such a stable region as ``plateau'', where the 
principle of minimal sensitivity is realized and the 
{\em exact} value will be reproduced.

\subsection{Properties of improved perturbative series}

\subsubsection{Details}

We will elucidate the concrete prescription of the improved 
perturbation method. 

The original perturbative series $F^N(\lambda, m)$ up to order $N$
is given in the following expression, where $f_n (m)$ is the $n$th 
order coefficient: 
\begin{equation}
	F^N (\lambda, m) 
	= 
	\sum_{n=0}^{N}\,\lambda^n\,f_n(m) \,.
\end{equation}
By the shift of the parameters, 
\begin{equation}
	\begin{aligned}
	\lambda & \longrightarrow g\,\lambda \,, \\
	m       & \longrightarrow m_0 + g (m - m_0) \,,
	\end{aligned}
\end{equation}
$F^N$ is deformed as,
\begin{equation}
	\begin{aligned}
	F^N (\lambda, m)
	\longrightarrow
	&
	\sum_{n=0}^{N}\,(g\lambda)^n\,f_n( m_0+g(m-m_0) ) \\
	= &
	\sum_{k=0}^{\infty}\,g^k\,
	\sum_{n=0}^{\min(N,k)}\,\lambda^n\,
	\frac{1}{(k-n)!}\,(m-m_0)^{k-n}\,f_n^{(k-n)}(m_0) \,,
	\end{aligned}
\label{eq:resummation}
\end{equation}
Here, each coefficient $f_n$ is expanded in Taylor series about $m=m_0$, 
with $f_n^{(k)}$ as $k$th derivative of $f_n(m)$ with respect to $m$.
The series is then reorganized in powers of $g$. Then we drop the ${\cal O}(g^{N+1})$ terms, 
and finally set $g$ to 1. 
As a result, we will have the improved series $\widetilde{F}^N$:
\begin{equation}
	\widetilde{F}^N 
	= 
	\sum_{n=0}^{N}\,\lambda^n\,
	\sum_{k=0}^{N-n}\,\frac{1}{k!}\,(m-m_0)^k\,f_n^{(k)}(m_0) \,.
\end{equation}

It turns out that the improved perturbation method replaces each 
coefficient of the original series, $f_n(m)$, by its Taylor 
expansion about a shifted point $m_0$,
\begin{equation}
	f_n(m) 
	\longrightarrow 
	\widetilde{f_n}(m; m_0) = 
	\sum_{k=0}^{N-n}\,\frac{1}{k!}\,(m-m_0)^k\,f_n^{(k)}(m_0) \,.
\label{eq:deformed_coeff}
\end{equation}
It should also be noted that the expansion in eq.~(\ref{eq:deformed_coeff}) 
is taken only up to $(N-n)$th order. 
This feature works to suppress the large fluctuations coming from 
higher order coefficients, which have caused divergent behavior of the 
original series. Thus we could expect better estimates by this 
improved perturbation method.

The important point is that the reorganized series should be 
truncated at finite order $N$. If we kept all the terms and 
if we could switch the order of two summations 
in eq.~(\ref{eq:resummation}), the procedure should become trivial, 
only to obtain the original series.

\subsubsection{Convergence}

The convergence property of the improved series with respect 
to $\lambda$ is, however, not affected significantly. 
The ratio of $(n+1)$th coefficient against $n$th one, 
\begin{equation}
	\frac{\widetilde{f}_{n+1}}{\widetilde{f}_{n}}
	=
	\frac{
	\sum_{k=0}^{N-n-1}\,\frac{1}{k!}\,(m-m_0)^k\,f_{n+1}^{(k)}(m_0)
	}{
	\sum_{k=0}^{N-n}\,\frac{1}{k!}\,(m-m_0)^k\,f_{n}^{(k)}(m_0)
	} \,,
\end{equation}
has no particular structures such as singularities, at least in 
obvious manner.	There, the parameters $m_0$ are determined 
according to the following argument; 
minimal sensitivity condition will be realized ideally when 
$p$th or lower derivatives of $\widetilde{F}^N$ with respect to the 
parameters $m_0$ become zero:
\begin{equation}
	\begin{aligned}
	0 
	&=
	\frac{d^p}{dm_0^{\phantom{0}p}}\widetilde{F}^N(\lambda, m; m_0) \\
	&=
	\sum_{n=0}^{N}\,\lambda^n\,
	\sum_{q=0}^{p-1}\,
		\begin{pmatrix}
			p\!\!-\!\!1 \\
			q
		\end{pmatrix} \,
		(-)^{q}\,
		\frac{1}{(N-n-q)!}\,(m-m_0)^{N-n-q}\,
		f_n^{(N-n+p-q-1)}(m_0) \,.
	\end{aligned}
\end{equation}
We will examine the condition further in later sections.

\subsubsection{Limitations}

As mentioned before, the improved perturbation method does not much alter 
the convergence properties. 
There are some class of functions such that the improved series 
actually is convergent in the ``plateau'' region, 
and the procedure of improved perturbation works as 
a sort of analytical continuation. 
For example, 
the function $\frac{1}{1+\lambda m}$ discussed 
in the previous subsection 
belongs to this class\cite{conv}. 
However, most of the series that appear in physics 
do not have the property like this. 
 
\begin{figure}
\subfigure[][]{%
	\includegraphics[scale=.8]{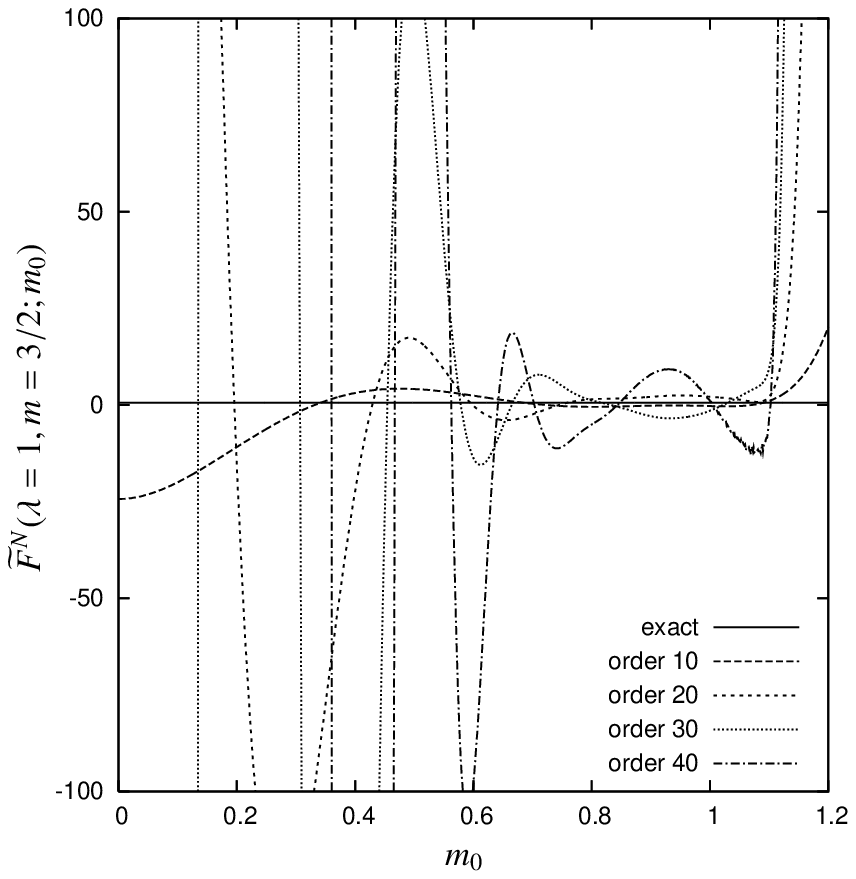}
\label{fig:3}
}
\hskip 2em
\subfigure[][]{%
	\includegraphics[scale=.75]{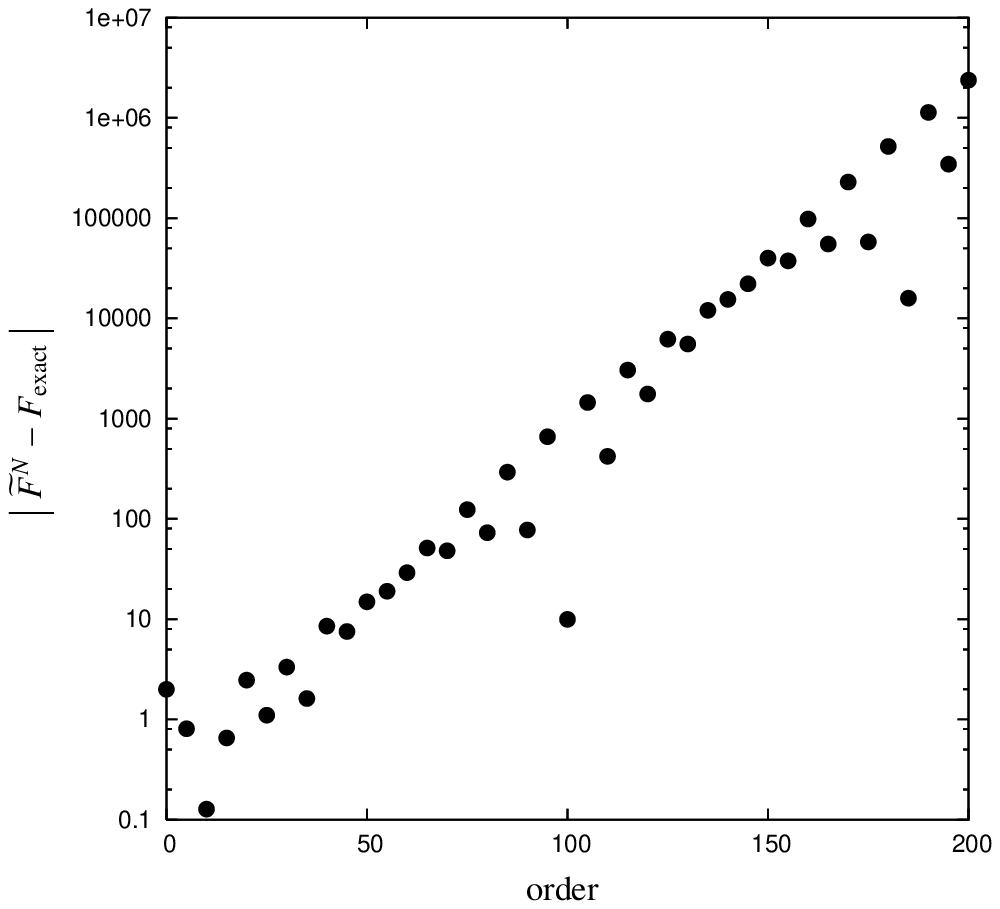}
\label{fig:4}
}
\caption{%
\subref{fig:3} Large fluctuation of $\tilde{F}$ as a function of $m_0$.
\subref{fig:4} Deviation between $\tilde{F}^N$ and the exact value. 
($m_0$ is chosen to be 0.95.)}
\label{fig:3and4}
\end{figure}

If the original series is asymptotic or is evaluated at outside 
of convergence region, the improved series 
may also show the divergent behavior at excessively high orders. 
We will illustrate this feature by a simple example. 
Consider an ordinary function $F$, 
\begin{equation}
	F(\lambda, m) 
	= \frac{1}{1 + \lambda m} 
	+ \frac{1}{1 + \lambda m + (\lambda m)^3} \,,
\end{equation} 
and generate a Taylor series about $\lambda=0$ up to order $N$. 
The radius of convergence of the original series is
$(\lambda m) \sim 0.7$.
Then, we apply the improved perturbation method to this series. 
As shown in Fig.~\ref{fig:3}, 
the improved series at $\lambda=1, m=3/2$ fluctuates violently with 
respect to $m_0$. 
%
%
The deviation from the {\em exact} value grows larger as the order 
increases. 
(The artificial parameter $m_0$ is chosen to be $m_0 = 0.95$.)
The improved perturbation method fails to reproduce the exact result in
this case at high enough orders (Fig.~\ref{fig:4}).

In order to obtain a reliable estimate by the improved perturbation method, 
we have to combine various information of the series of different 
orders, and examine synthetically the whole profile of the series.

\subsubsection{Ring properties}

In this section, we examine how the improved perturbation method affects 
the function ring $C^\infty$. 
We are especially interested in the derivative operation, for 
the expectation value of an operator is related to the derivative 
of the free energy with the corresponding source term.

Usually we can introduce an equivalence relation into a set of 
elements of function ring, and classify them into equivalence classes. 
Now we insist that the plateau criteria works for such 
classification; two elements of the function ring are equivalent 
if there exists intersection between ``plateau'' of each function 
where minimal sensitivity is realized.

First we consider the add operation. 
Assume that we have two functions, $F$ and $G$, and the improved 
series, $\widetilde{F}^N$ and $\widetilde{G}^N$, respectively. 
The procedure to obtain the improved series is basically Taylor 
expansion, and thus linear. 
If $\widetilde{F}^N$ and $\widetilde{G}^N$ have intersecting region where 
minimal sensitivity condition for each series is realized 
simultaneously, then the improved series of sum of the functions, 
$\widetilde{(F\!+\!G)}^N$ also stays stable and reproduces the {\em exact} 
value there. 

On the other hand, when $\widetilde{F}^N$ and $\widetilde{G}^N$ do not have overlapped parameter 
region, the sum $\widetilde{(F\!+\!G)}^N$ does not in general bear stable region in their plateaux.

Thus, the improved perturbation method preserves the add operation among 
the elements of function ring that belong to the same equivalence 
class based on the presence of intersection of ``plateau''.

Unlike the case of the add operation, there is no reliable 
discussion on the product operation of function ring. 
However, from several examples, it seems that the procedure 
also preserves the product operation among the elements of the 
same equivalence class.

Next we consider the derivative operation. 
Similar to the ordinary function, the derivative of the improved 
series, $\widetilde{F}^N(\lambda,m; m_0)$, with respect to $m$ 
will be defined as follows:
\begin{equation}
	\frac{d}{dm} \widetilde{F}^N(\lambda, m; m_0)
	= 
	\lim_{\epsilon\rightarrow 0}\,
	\frac{\widetilde{F}^N(\lambda, m+\epsilon; m_0) 
	- \widetilde{F}^N(\lambda, m; m_0)}{\epsilon} \,.
\label{eq:improved_deriv}
\end{equation}
We will designate as $\mbox{R}(\lambda, m)$ the region in 
the parameter space $m_0$ where the minimal sensitivity condition 
is satisfied. 
We also call it as ``region of minimal sensitivity'' below.

If two regions of minimal sensitivity for physical parameters at 
$m+\epsilon$ and $m$ have overlap, 
\begin{equation}
	\mbox{R}(\lambda, m+\epsilon)\,
	\cap\,
	\mbox{R}(\lambda, m) \neq \emptyset \,,
\end{equation}
we would conclude from the argument for the add operation, that 
the derivative function defined in eq.~(\ref{eq:improved_deriv}) 
bears the region of minimal sensitivity in the intersection of 
the two regions. 
In short, for a function and its derivative, 
the region of minimal sensitivity will be expected to emerge 
somehow close to each other.

Otherwise, it may happen that the region of minimal sensitivity 
of $\widetilde{F}^N$ and its derivatives are uncorrelated. 
Then, in what situation does the overlap vanish, {\it i.e.}
$\mbox{R}(\lambda, m+\epsilon) \cap \mbox{R}(\lambda, m) = \emptyset$ ?
This question is deeply related to the critical behavior of the 
underlying physical model. 
In usual cases, free energy and various expectation values are 
regular with respect to moduli parameters, and the region 
$\mbox{R}(\lambda, m)$ evolves regularly as well. 
However, at singular points of moduli space such as those where 
the symmetry of the model is enhanced, 
$\mbox{R}(\lambda, m)$ should emerge in a quite different manner; 
thus even for infinitesimally small deviation of moduli parameter, 
$\mbox{R}(\lambda, m)$ and $\mbox{R}(\lambda, m+\epsilon)$ 
may not have overlaps.

Such a situation often occurs in the physical models of our interests.
For example, in Ising model, the region of minimal sensitivity 
drastically changes between high temperature phase and low temperature 
phase. 
In the IIB matrix model (which will be examined extensively in later 
sections), 
we are interested in the massless region, where the parameters 
become singular. 
In these cases, the free energy and the expectation values of 
various observables show different behavior, and therefore 
we have to treat them separately.

\section{Plateau conditions\label{sec:plateau}}

In this section, we will elucidate the criteria how to determine 
desirable regions of artificial parameters, 
and to estimate the approximate value from 
the improved series.
A guiding principle, minimal sensitivity, will be realized in 
regions in the parameter space where the improved function 
stays rather stable. 
We call such region as {\em plateau}.

The subject of this section is to develop a concrete and objective 
scheme for identifying {\em plateau}.
Besides, the notion, ``stable'' is obscure, and so the procedure may 
also provide an inductive definition of {\em plateau} itself.

\subsection{Minimal sensitivity and plateau}

The improved perturbation method introduces artificial parameters, $m_0$, 
which we need to determine by some means.
Here we adopt as a guiding principle, {\em minimal sensitivity}, 
that the improved function should depend least on those parameters; 
this is because $m_0$ were originally introduced as the nominal 
shift of parameters, though the dependence on them appeared 
due to dropping the ${\cal O}(g^{N+1})$ terms. 
If there exists a region in the parameter space $m_0$ where the 
dependence on $m_0$ vanishes effectively, the {\em exact} value should 
be reproduced there. In the following we denote simply by
$\widetilde{F}^N$ the improved series 
$\widetilde{F}^N(\lambda=\lambda_*,m=m_*;m_0)$ where we substitute each
desirable values $\lambda_*$ and $m_*$ for $\lambda$ and $m$. This
series is function of artificial parameters $m_0$.  

The principle of minimal sensitivity is realized through the 
emergence of the region in the parameter space 
where the improved series stays stable against any variation 
of artificial parameters; we call such a region as ``plateau''. 
What we have to do for determining $m_0$ is to identify where 
the improved series becomes stable.

When the number of parameters are one (or at most two), the 
stable region of the function will be pointed out by just 
drawing a graph of it. 
However, the visualization will not be possible for the cases 
with more than two parameters. 
Moreover, even though it is possible, it may sometimes lead 
to misjudgment by changing the scale arbitrarily. 
Therefore we need a concrete scheme for identifying the plateau 
applicable to multi-parameter space, which is not affected 
by intuitive guesses. 

\subsection{Identifying plateau}

The ideal realization of the principle of minimal sensitivity 
may have such a property that the improved series is totally independent 
of the artificial parameters $m_0$ in a region. 
It involves the situation that 
all orders of derivatives of the improved series 
with respect to $m_0$ are zero in the region. 
However, such an ideal ``plateau'' is not realized in the actual cases 
because we have only finite order of series. 
Typical profile of the improved series that forms plateau 
exhibits a flat region in which the series fluctuates bit by bit; 
it would accompany a number of local maxima and minima there. 
Thus, to turn the argument around, we consider the accumulations 
of extrema as indications of plateau (or its candidates).

It should be noted that there is also a case when the series 
becomes stable without forming any extrema in that region; 
we may miss such plateau by the above speculation. 
We will discuss this type of plateau in later section.

\subsubsection{Cluster identification}

Assume that we have already found extrema of the improved series 
in the (multi-dimensional) parameter space. 
Next issue is to identify the accumulation among them.
We denote the coordinate of $i$th extrema by $\vec{x}_i$, and 
the distance between $i$th and $j$th extrema by $d_{ij}$. 
The definition of distance in the parameter space will 
be presented later. 

The distribution $\rho(r)$ of the distances $d_{ij}$ shows 
characteristic behavior according to the presence of clusters 
of points.
Here, $\rho(r)$ is given by the number of pairs $(i,j)$ whose 
distance $d_{ij}$ falls on to $r < d_{ij} < r+\delta r$.
%
%
\begin{figure}
\begin{center}
\subfigure[][]{%
	\begin{tabular}{c}
	\includegraphics[scale=.6]{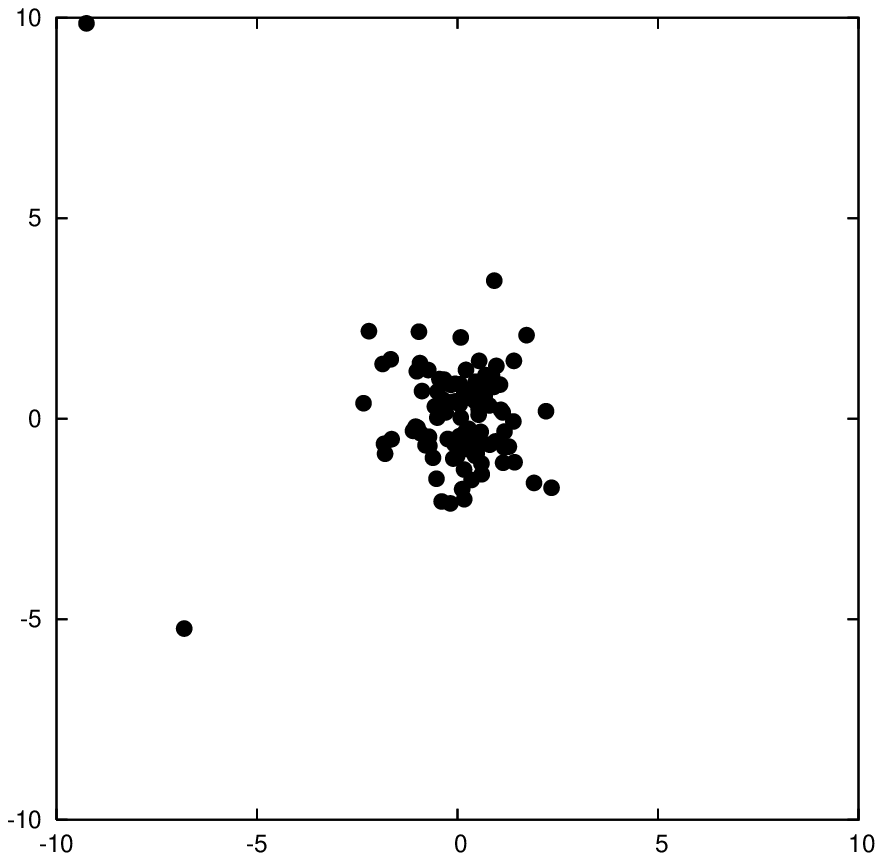} \\
	\includegraphics[scale=.6]{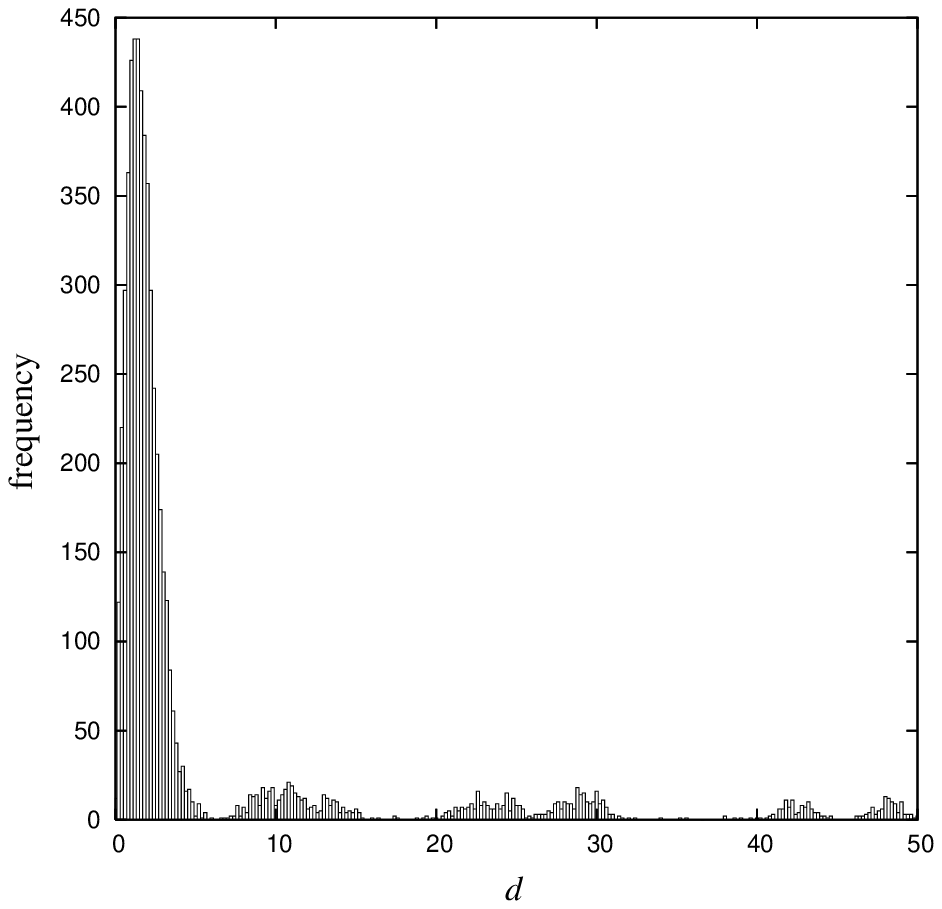} \\
	\end{tabular}
\label{fig:cluster:a}
}
\hskip 3em
\subfigure[][]{%
	\begin{tabular}{c}
	\includegraphics[scale=.6]{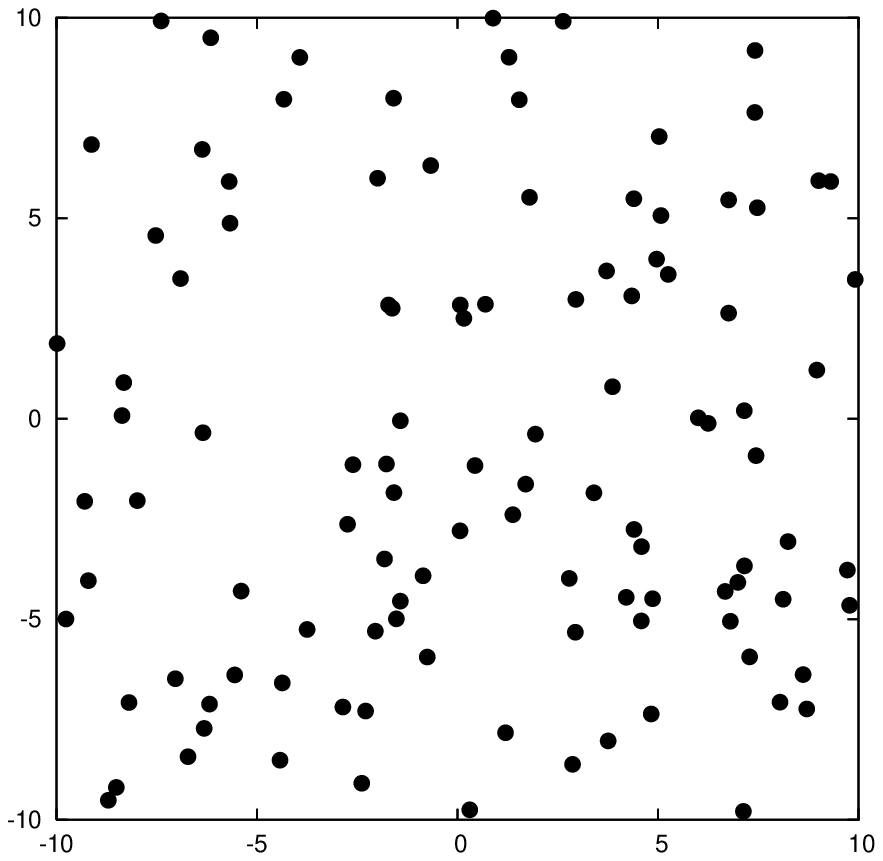} \\
	\includegraphics[scale=.6]{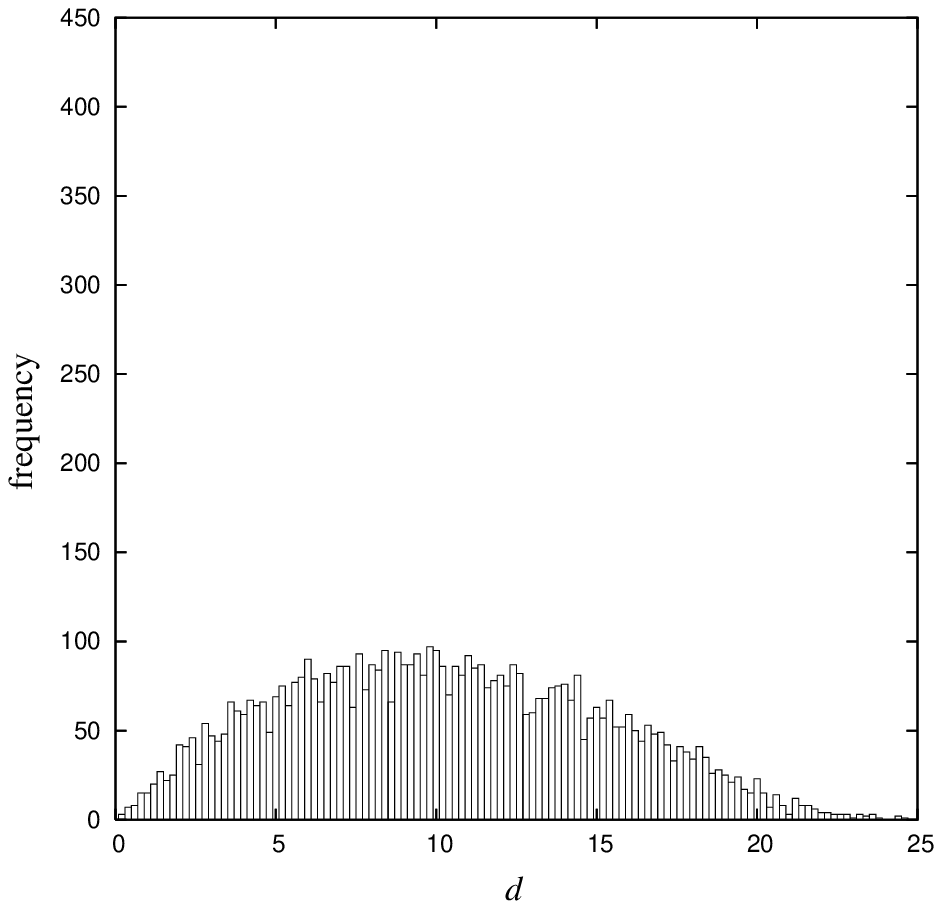} \\
	\end{tabular}
\label{fig:cluster:b}
}
\end{center}
\caption{%
Distributed points (above) and distributions of distance 
between points (below) for each case: 
\subref{fig:cluster:a} forming the cluster, and 
\subref{fig:cluster:b} straggling uniformly.
}
\label{fig:cluster}

\end{figure}
As depicted in Fig.~\ref{fig:cluster}, when one or more clusters of points 
exist, $\rho(r)$ has a large peak near $r=0$ and some small 
bumps at larger $r$. 
Let $n$ the number of points in the cluster. Then the peak height 
is proportional to $n^2$, while those of other bumps are of order $n$.
If there are no such clusters and the points are scattered 
rather uniformly, $\rho(r)$ shows continuous distribution. 
This concept can be extended straightforwardly to the case 
in which more than one cluster are formed. 
From those distinctive properties, we are able to identify the 
formation of clusters and the set of points belonging to them. 

There is no intrinsic definition of distance $d_{ij}$ in 
multi-dimensional parameter space. Therefore we have to 
choose a suitable definition. It introduces metric space 
and must obey the triangle inequality.

A na\"ive choice is as follows:
\begin{equation}
	d_{ij}^{\phantom{ij}2} 
	= 
	\bigl| \vec{x}_i - \vec{x}_j \bigr|^{2} \,.
\end{equation}

As another choice, if we can define a reasonable metric $g_{ab}$ 
from some argument such as dimensional analysis, we will have,
\begin{equation}
	d_{ij} = \int_{\text{geodesic}}\, \sqrt{-g}\,ds \,,
\end{equation}
along geodesic between $\vec{x}_i$ and $\vec{x}_j$.
The geodesic equation may not be solved nor have any solution;
we would alternatively choose a linear interpolation with weights 
by metric as,
\begin{equation}
	d_{ij} 
	= \sum_k\,\sqrt{-g(t_k)}\,
	\bigl| \vec{x}(t_k)  -\vec{x}(t_{k+1}) \bigr| \,,
	\quad
	\vec{x}(t_k) = \vec{x}_i + \frac{k}{N}(\vec{x}_j - \vec{x}_i) \,, 
\end{equation}
though this $d_{ij}$ may not satisfy triangle inequality.

It is not always that two extrema which close to each other in the meaning of the above
distances belong to one stable region. This is because it is possible
that between these extrema improved series varies its value
drastically and vicinity of two extrema encounters by chance. In order
to avert this situation, we consider one other choice of distance.   
This choice comes from extending the concept of moduli space.
By the definition of distance we intend to express the degree 
to which two of the extrema of a function $\widetilde{F}^N$ resemble 
each other. 
Thus we include the deviation of $\widetilde{F}^N$ and its derivatives 
at these two points into the notion of distance:
\begin{equation}
	d_{ij}^{\phantom{ij}2} 
	= \bigr| \vec{x}_i - \vec{x}_j \bigl|^2
	+ \sum_k\,w_k\,
	\Biggl|\,\,
		\frac{d^k\widetilde{F}^N}{dx^k} \Bigl|_{\vec{x}_i}
		-\frac{d^k\widetilde{F}^N}{dx^k} \Bigl|_{\vec{x}_j}
	\Biggr|^2 \,.
\end{equation}
Moreover we can also include the transition as orders of perturbation increase:
\begin{equation}
	d_{ij}^{\phantom{ij}2} 
	= \bigr| \vec{x}_i - \vec{x}_j \bigl|^2
	+ \sum_{M,M'=0}^{N} \omega_{MM'} \sum_k\,w_k\,
	\Biggl|\,\,
		\frac{d^k\widetilde{F}^M}{dx^k} \Bigl|_{\vec{x}_i}
		-\frac{d^k\widetilde{F}^{M'}}{dx^k} \Bigl|_{\vec{x}_j}
	\Biggr|^2 \,.
\end{equation}
The relative weights $w_k$ and $\omega_{MM'}$ above should be chosen according to 
the models considered.

In general, we can not introduce a particular choice of metric
naturally. Then we try some definitions of metric to a model and
investigate the improved perturbative series carefully.

\subsubsection{Weighted histogram analysis}

If the improved series $\widetilde{F}^N$ bears a plateau, 
the distribution of the improved series in a region enclosing 
the plateau should form a peak corresponding to the value of 
$\widetilde{F}^N$, 
which gives an estimate of the function on the plateau. 

Therefore, we choose a region (for ease of operation, we usually 
consider rectangular one) which encloses accumulation of extrema, 
and evaluate distribution $\rho(F)$ of $\widetilde{F}^N$ \cite{Nishimura:2002va},
\begin{equation}
	\rho(F) = \int_{A} dx\,1 \,,
	\quad
	A = \{ x\,\bigl|\, F < \widetilde{F}^N(x) < F\!+\!\delta F \} \,.
\end{equation}

In multi-dimensional parameter case, the extent of the plateau is 
not always large nor the shape isometric in the parameter space. 
Then we need a method to extract the estimated value of $\widetilde{F}^N$ 
on plateau effectively 
even when the region to be examined is taken roughly.

The plateau is, if ideally realized, characterized by the feature 
that the first and higher derivatives of $\widetilde{F}^N$ should 
be zero or small.
To enhance the contribution from such flat region, we introduce 
weight function to the distribution as, 
\begin{equation}
	\rho(F) = \int_{A}dx\,w \,.
\end{equation}
There may be some choices of the weight function $w$; for example,
\begin{align}
	w_1 
	&= 
	\frac{1}{
		\Bigl|\,\sum_i \frac{d\widetilde{F}^N}{dx_i}\,\Bigr|^2
	} \,, 
\label{eq:1std} \\
	w_2 
	&= 
	\frac{1}{
		\Bigl|\,\det\frac{d^2\widetilde{F}^N}{ dx_i dx_j}\,\Bigr|^2
	} \,,
\label{eq:2ndd}
\end{align}
and so forth. 
Here we denote the $i$th component of coordinates by $x_i$. 

The histogram of weighted distribution shows sharp peak corresponding 
to the zeros of $k$th derivatives of the improved series. 
Even though the region in question is taken roughly, the contribution 
from the fluctuating part of $\widetilde{F}^N$ is well suppressed, 
and we will have an estimate of $\widetilde{F}^N$ on the plateau.

There is also another choice of weight function based on the 
convergent behavior of the series on the plateau. 
If the improved series converges to some value on the 
plateau, it implies that the difference between the improved 
series of order $N$ and order $N+1$ diminishes. 
To reflect such speculation, we would better choose a weight function 
by 
\begin{equation}
	w_3 = \frac{1}{\Bigl|\,\widetilde{F}^{N+1} - \widetilde{F}^N\,\Bigr|^2} \,.
\end{equation}

In general, we combine the above weight functions on a case-by-case
basis. In the following investigation into the IIB matrix model, we use
the two weight functions $w_1$ (\ref{eq:1std}) and $w_2$ (\ref{eq:2ndd}).

\subsubsection{Signal and noise}

When the number of parameters are one (or at most two), we can 
easily recognize the flat region from the graph. 
However, there is no such simple scheme in multi-parameter case 
in general. 
We are trying to identify plateau from the accumulation of 
the extrema as a clue.
The information we have so far for the improved series 
$\widetilde{F}^N$ is just the location of extrema and profile of 
(weighted) distribution of $\widetilde{F}^N$ in some specified regions.
The accumulations of extrema provides candidates of the plateau. 
Among these candidates we have to distinguish the true plateaux, though  
yet we do not have definite argument on this subject. 
We exemplify one typical false signal, ``overshoot''.

\begin{figure}[t]
\includegraphics[scale=.8]{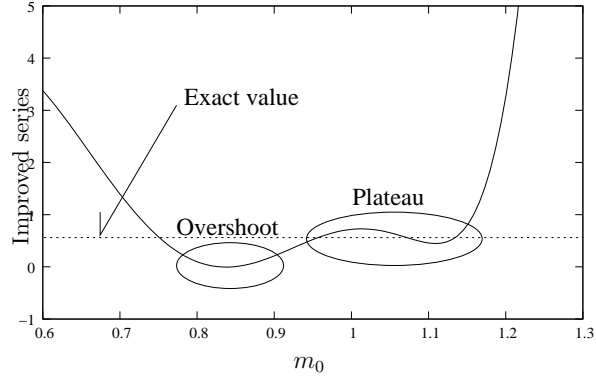}
\caption{A characteristic profile of overshoot and plateau.}
\label{fig:overshoot-1}
\end{figure}

It often occurs when a discrete function is approximated 
by a series of polynomial 
(consider, for example,  Fourier transform of rectangular waves) 
that the series deviates relevantly at discontinuities, 
which remains with even higher orders taken into account. 
Such behavior is called ``overshoot''.
In the current case, the improved series turns to be an 
approximation of a constant function by a set of polynomials 
because the exact value is totally independent of artificial parameters. 
Therefore the improved series are apt to show typical overshoot
behavior as well (Fig.~\ref{fig:overshoot-1}).

\begin{figure}[t]
\begin{center}
\subfigure[][]{%
\includegraphics[scale=.6]{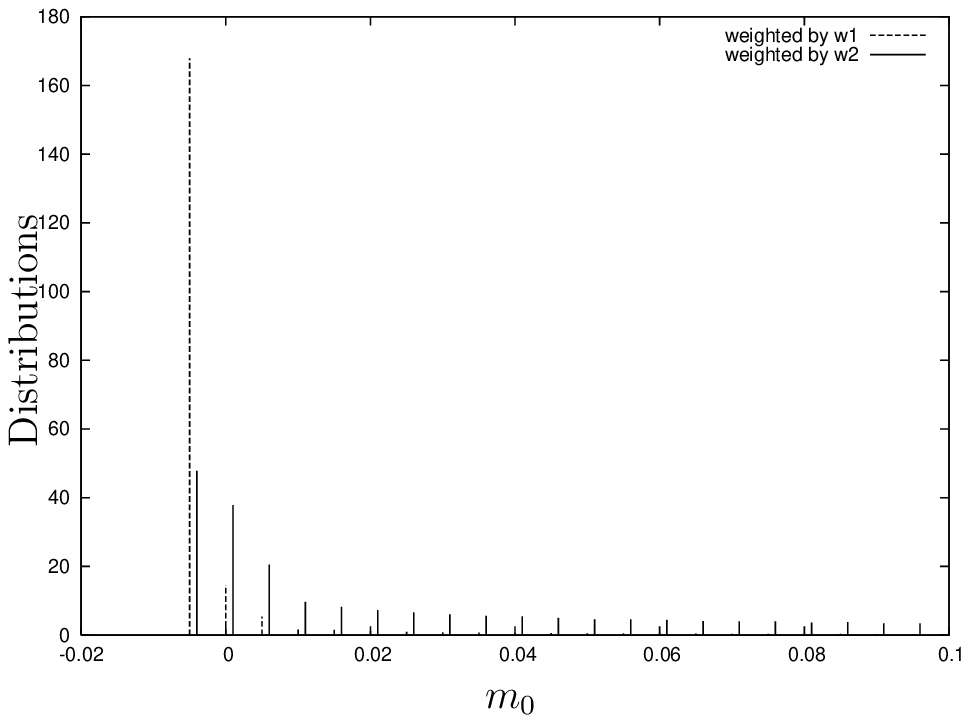}
\label{fig:overshoot-2}
}
\hskip 4em
\subfigure[][]{%
\includegraphics[scale=.6]{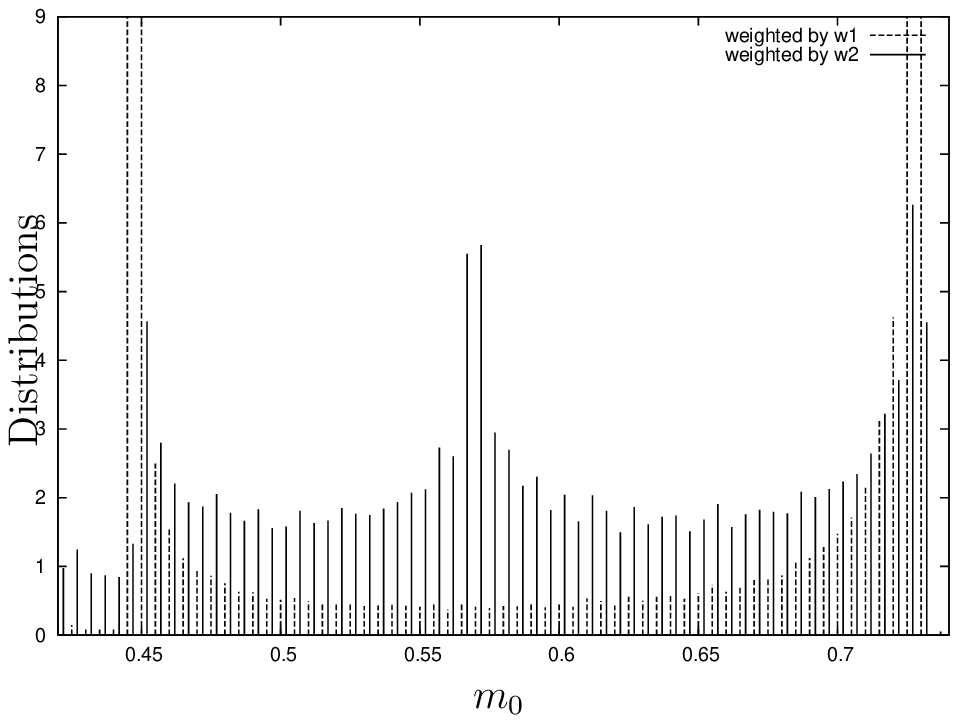}
\label{fig:overshoot-3}
}
\end{center}
\caption{(a):Distribution weighted by $w_1,w_2$ in overshoot region and 
(b)distribution weighted by $w_1,w_2$ in plateau region.}
\end{figure}

The overshoot is rather isolated from other extrema that 
belong to a plateau; this should be reflected in the 
characteristic profile of weighted distributions. 
The overshoot gives minimum (or maximum) in its neighbor, so 
the histogram of the distribution weighted by $w_1$ shows sharp peak, 
and it is cut away on lower (or upper) side (Fig.~\ref{fig:overshoot-2}). 
Moreover it often happens that the peaks in the distributions with second
and higher derivatives deviate from peaks in that with the first derivative.
It is because the 
overshoot appears as isolated extrema away from those forming 
flat region. On the contrary, in the case of plateau, the histogram of
the distribution weighted by $w_1$ and $w_2$ shows symmetric shape and
a peak of the distribution weighted by $w_2$ lies between (or ideally
on) peaks of the distribution weighted by $w_1$
(Fig.~\ref{fig:overshoot-3}). 
For the reason stated above, 
a cluster of extrema which corresponds to overshoot can be
distinguished from plateau by investigating the shape of distributions.
%
%

Besides, it seems that the overshoots of different orders 
tend to form a line in the parameter space. 
If there is such sequential structure, we had better suspect 
that they are overshoots.

\subsection{Pitfalls and special cases}

\subsubsection{Smooth plateau}

\begin{figure}
\includegraphics[scale=.8]{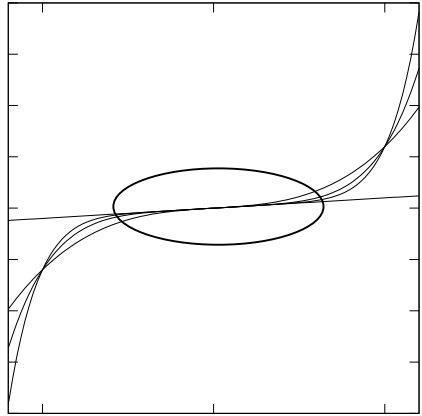}
\caption{A conceptual profile of {\em smooth} plateau.}
\label{fig:gentle_plateau}
\end{figure}

In the prescription thus far explained, we assumed that the 
improved series $\widetilde{F}^N$ shows fluctuating behavior on 
the plateau, accompanied by the accumulation of extrema.
If there is a region where $\widetilde{F}^N$ becomes stable but 
varies gently without forming extrema (Fig.~\ref{fig:gentle_plateau}), it should also be considered 
as plateau, though the above procedure based on the existence of extrema
will not be applicable to such cases. We call this type of plateau as 
``smooth plateau''. 
However, the histogram will show the peaked structure corresponding 
to this smooth plateau. 
The weighted distribution should provide a reasonable estimate 
of the value.
%
%

The problem is that we do not have any guide to locate the 
position of plateau in this case. 
One clue is the consistency with the plateau of other 
physical quantities. 
Though the improved series of different observables may show 
distinctive behavior, the plateau corresponding to the physical 
states should somehow be realized in each series.
We have to guess the region of smooth plateau in parameter space 
based on those information. 
If we find a plateau in analysis of an observable, we must suspect 
the appearance of smooth plateau in neighborhood of that region 
in the case of the other observables even for absence of any extremum.  

It should be noted that we have to take care not to be confused 
with mere asymptotic behavior of the series. 
A concrete example will be shown in the analysis 
of the IIB matrix model in later sections.

\subsubsection{Plateau on special hypersurface}

Consider a hypersurface in the parameter space where one of 
the parameters is zero. 
If we change the coordinate from $m_0$ to $x_0 = \frac{1}{m_0}$, 
the first derivative with respect to the new coordinate 
becomes, 
\begin{equation}
	\frac{d\widetilde{F}^N}{dx_0} = - m_0^{\phantom{0}2}\,\frac{d\widetilde{F}^N}{dm_0} \,.
\end{equation}
Even though $\frac{d\widetilde{F}^N}{dm_0}$ does not vanish, $\frac{d\widetilde{F}^N}{dx_0}$ 
becomes zero at $m_0 = 0$.
Thus we have to investigate the case separately when some of 
parameters take zero or $\infty$. This is because we do not know what
choice of the artificial parameters is preferred.

Those cases are significant in that $m_0 = 0$ or $\infty$ 
occasionally corresponds to the point where symmetry of the original 
physical model enhances; therefore the improved series may have 
singular structure.

\subsubsection{Multiple plateaux}

In the multi-parameter case, it is likely that there exists more 
than one plateau in the parameter space. 
If each of them can be interpreted as a physical state, some 
correspond to stable vacua, while others to the unstable ones. 
This situation is realized in Ising model~\cite{Aoyama:2005nd}. 
When we carry out the improved perturbation for the IIB matrix model, we 
expect that multiple plateaux are realized corresponding respectively to 
different vacua with different symmetry. For this reason we must pick up all candidates of
plateau and examine these carefully.

\subsection{Short summary}

In this section, we discussed the concept of ``plateau'' and the 
prescription to identify them in the improved series. This improved 
perturbation method must be applied independently 
for individual observables. 
   
To briefly summarize the procedure for each observable, we first find extrema of 
the series as a function of artificial parameters $m_0$, and then identify 
the accumulations of the extrema which should be considered as 
candidates of plateau.
Next we evaluate weighted distributions in the region which 
encloses each of the candidates, 
and distinguish the plateau among them. 
The histograms also provide estimates of the {\em exact} value 
from the improved series. 

This procedure assumes typical profile of the improved function. 
It will not be applicable when the function stays stable without 
forming extrema. 
In these cases, we have to locate the flat region referring to 
the information of other physical quantities, such as symmetry 
considerations.
We also have to take care of the special hypersurface where 
some of parameters take zero or infinity; such cases may be 
relevant in the original physical models.

\section{IIB matrix model\label{sec:ikkt}}

In this section we apply the improved perturbation method to 
the IIB matrix model and investigate its non-perturbative solutions. 
In particular we concentrate 
on the free energy and the second moment of eigenvalue distribution. 
By comparing the values of free energy of the solutions, we can see 
which solution is most preferred. 
In the IIB matrix model, the distribution of eigenvalues is interpreted 
as the space-time itself. 
Therefore we can recognize the shape of the universe 
by computing the eigenvalue distribution. 
To be more precise, 
the square root of the second moment of eigenvalues distribution 
gives the scale of extent to each direction of universe. 
The deviation of the ratio of space-time extent between directions 
from unity indicates the degree of anisotropy of the universe. 

\subsection{Model setup}

The action of the IIB matrix model is 
\begin{equation}
	S_{\text{0D-YM}} = 
	\Tr\left(
		-\frac{g_{\rm YM}^2}{4} \left[ X_a, X_b \right]^2
		-\frac{g_{\rm YM}}{2} \bar{\Psi} 
		\left[ X_a, \Gamma^a \Psi \right]
	\right) \,,
\end{equation}
where $X_a$ $(a=1 \dots 10)$ and 
$\Psi^{\alpha}$ $(\alpha=1 \dots 16)$ are all 
$N \times N$ hermitian matrices which belong to a vector and 
a Majorana-Weyl spinor representation of $\SO(10)$, respectively. 
This action can be 
obtained by the dimensional reduction from ten-dimensional $\U(N)$ 
supersymmetric Yang-Mills action 
to zero dimension. 
It has the symmetry of all global $\SO(10)$, $\U(N)$ 
and type IIB supersymmetry. 

We perform the following change of scales: 
\begin{equation}
\begin{aligned}
	X_a &\longrightarrow N^{\frac{1}{4}}X_a \,, \\
	\Psi&\longrightarrow N^{\frac{1}{8}}\Psi \,, \\
	g_{YM}^2 N &\longrightarrow \lambda \,.
\end{aligned}
\end{equation}
Then the action takes the form: 
\begin{equation}
	S = N \Tr \left(
		-\frac{1}{4} \lambda \left[ X_a, X_b \right]^2 
		-\frac{1}{2} \sqrt{\lambda} 
		\bar{\Psi} \left[ X_a, \Gamma^a \Psi \right]
	\right) \,.
\end{equation}
In order to carry out the perturbative expansion, 
we add the following propagator terms to the action 
\begin{equation}
	S_0 = 
	N \Tr \left(
		\frac{1}{2}\sum_{a,b=1}^{10} m_{ab} X_a X_b 
		+\frac{1}{2}\sum_{a,b,c=1}^{10} 
		m_{abc} \bar{\Psi} \Gamma^{abc} \Psi 
	\right) \,.
\label{eq:massterm}
\end{equation}
They are the most general quadratic terms. 
The fermionic bi-spinors are expanded by antisymmetric tensors 
by using gamma matrices. 
In the case of ten-dimensional Majorana-Weyl spinor, 
by considering a chirality, only gamma matrices of rank three 
are allowed. 

The scaling behavior of the Feynman rules for the amplitude 
with respect to $N$ and $\lambda$ is given as follows. 
\begin{center}
\begin{tabular}{ll}
propagator for bosonic fields $X_a$:    
& \hskip3em $\displaystyle \sim \ \frac{1}{N}$, \\
propagator for fermionic fields $\Psi$: 
& \hskip3em $\displaystyle \sim \ \frac{1}{N}$, \\
3-point vertex: 
& \hskip3em $\displaystyle \sim \ N \sqrt{\lambda}$, \\
4-point vertex: 
& \hskip3em $\displaystyle \sim \ N \lambda$. \\
\end{tabular}
\end{center}
If we denote the genus of dual surface by $\mathfrak{g}$
and the number of faces of dual surface by $N_F$, 
we find that the contribution of a single diagram is 
\begin{equation}
	\text{amplitude} \sim N^{2(1-\mathfrak{g})} \lambda^{\frac{N_F}{2}} \,.
\end{equation}
If we take the limit $N \to \infty$ with $\lambda$ fixed, 
the planar diagrams ($\mathfrak{g}=0$) give the leading contributions. 
In this case the amplitude is given by 
\begin{equation}
	\text{amplitude} \sim N^2 \sum_{n}\lambda^n f^n \,. 
\end{equation}
Free energy $F$ is evaluated by the sum of 
connected bubble diagrams in terms of propagators $m_a$ and $m_{abc}$ as 
\begin{equation}
	\frac{F}{N^2} = -\ln\left(\frac{1}{m}\right) + \cdots \,.
\end{equation}
The expectation value of the second momentum of eigenvalue distributions 
is obtained by 
the derivatives of free energy with respect to the parameters.
\begin{equation}
\begin{aligned}
	\frac{1}{N^2} \frac{\partial F}{\partial m_a} 
	&= 
	\left\langle \frac{1}{N} \Tr X_a^2 \right\rangle \\
	&= 
	\frac{\partial}{\partial m_a} 
	\left( -\ln\left(\frac{1}{m}\right) + \cdots \right) \,.
\end{aligned}
\end{equation}

The original IIB matrix model does not have the propagator term 
eq.~(\ref{eq:massterm}). Therefore we evaluate the observables at 
$m_a=m_{abc}=0$. They are singular in this region of parameters 
in the above perturbation. We use the improved perturbation to 
extrapolate to the massless case. 

In general, all parameters $m_{ab}$, $m_{abc}$ have to be 
taken into account. 
We may perform $\SO(10)$ rotation to reduce $m_{ab}$ 
to the diagonal form at the stage of action. 
We will impose further restrictions from the practical reason 
in the later sections. 

Before going into a concrete computation, we make some assumptions. 
In computing the free energy we have to evaluate 
all connected bubble diagrams. 
As we proceed to the higher order of perturbation, the number of
diagrams becomes enormous. 
We consider such a situation that 
the $\U(N)$ symmetry stays intact. Throughout this article 
we concentrate on the spontaneous breakdown of Lorentz symmetry\footnote{%
The breakdown of $\U(N)$ gauge symmetry will be discussed in the 
subsequent paper \cite{cs}.
}. 
Then we can omit the tadpole amplitude 
\begin{equation}
	\langle (X^a)_{ij} \rangle =0 
\end{equation}
for the matrix corresponding to the adjoint representation of $\SU(N)$ 
subalgebra. 
By the argument of the gauge symmetry alone we can not forbid 
the tadpole amplitude of $\U(1)$ part. 
If it takes non-zero expectation value $x^a$, 
we perform the field redefinition 
\begin{equation}
	X^a \rightarrow X^a + \frac{x^a}{N} \mathbf{1}_{N\times N} \,.
\label{eq:u1trans}
\end{equation}
By this redefinition we can set the tadpole amplitude to zero, 
\begin{equation}
	\langle \Tr(X^a) \rangle = 0 \,.
\end{equation}
As we show later the effective action bears dynamical mass term. Then 
the zero point of free energy is shifted 
due to the redefinition of fields by eq.~(\ref{eq:u1trans}). 
Thus we can not use the zero of free energy as a characteristics 
of supersymmetry. 

Under the assumption above, 
the sum of connected bubble diagrams reduces to 
the sum of one-particle irreducible (1PI) bubble diagrams. 
Thus far we have to evaluate 1PI bubble diagrams in order to obtain 
free energy. 

In the estimation of 1PI bubble diagrams we can drastically simplify 
the computation by using 2PI free energy 
\cite{Kawai:2002jk}. 
At first we compute 2PI graph amplitude $G_{\rm 2PI}$ by using full 
propagators 
\begin{equation}
\begin{aligned}
	\langle (X_a)_{ij} (X_b)_{kl} \rangle 
	&= 
	\frac{1}{N} c_{ab} \delta_{il} \delta_{jk} \,, \\
	\langle (\Psi_{\alpha})_{ij} (\Psi_{\beta})_{kl} \rangle 
	&=
	\frac{1}{N} \frac{1}{3!} u_{abc} ({\cal C}\Gamma^{abc})_{\alpha \beta} 
	\delta_{il}\delta_{jk} \,,
\end{aligned}
\end{equation}
in place of perturbative propagators $m_{a}, m_{abc}$. 
Here we use \textit{double line} notation, 
and $\cal C$ is the charge conjugation matrix. 

From $G_{\rm 2PI}$ we obtain 1PI bubble amplitude $F_{\rm 1PI}$ by 
Legendre transformation \cite{Fukuda:1995im}\footnote{%
For details see also Ref.\cite{Kawai:2002jk}.
}.
\begin{gather}
	\mathfrak{F}(m_a, m_{abc}; c_a, u_{abc})
	\equiv 
	G_{\rm 2PI}(m_a, m_{abc}) 
	+ \frac{1}{2} \sum_{a=1}^{10} m_a c_a
	- \frac{1}{2} \sum_{a,b,c=1}^{10} m_{abc} u_{abc} 
	\,, \\
	0
	=
	\left.\frac{\partial \mathfrak{F}}{\partial m_a}\right|_{c_a=\bar{c}(m)} \,, 
\label{eq:SD} \\
	F_{\rm 1PI}
	=
	\left.\mathfrak{F}\right|_{c_a=\bar{c}(m)} \,.
\label{eq:legendre} 
\end{gather}
Eq.~(\ref{eq:SD}) corresponds to the Schwinger-Dyson equation which
relates full propagators $c_a$ and $u_{abc}$ to perturbative propagators 
$m_a$ and $m_{abc}$, and 
eq.~(\ref{eq:legendre}) produce the 1PI amplitude using bare propagators 
$m_a$, $m_{abc}$.

In the present article, we use the result of the perturbative expansion 
up to eighth order in $\lambda$ of Ref.~\cite{Aoyama:2006rk}.

\subsection{Ansatz}

In this subsection, we make assumptions for the pattern of 
symmetry breakdown. 
In the case of the IIB matrix model, the total number of 
artificial parameters is quite large, namely, 
10 real numbers for $m_{ab}$ (assumed to be diagonalized by 
$\SO(10)$ rotation), and 120 for $m_{abc}$. 
It will demand an enormous effort to search for plateau 
in this vast space of parameters. 
Therefore, we impose restrictions on the configuration by 
considering symmetry to diminish the number. 
In concrete manner, we make same assumptions for the full
propagators $c_{ab}$ and $u_{abc}$ because we compute 2PI bubble
diagrams by using these parameters.  

In the former works%
\cite{Nishimura:2001sx,Kawai:2002jk,Kawai:2002ub,Aoyama:2006rk}, 
the configurations called $\SO(d)$ ansatz 
have been intensively examined which preserve $\SO(d)$ 
rotational symmetry. 
The guideline of choice is described as follows. 
First, $\SO(d)$ subgroup of $\SO(10)$ is chosen 
to which directions the expectation values of fermionic two-point 
function $u_{abc}$ are zero. $d$ is taken from 1 to 9. 
Toward the rest of the directions $u_{abc}$ may have non-zero value. 
Since $u_{abc}$ is a rank three anti-symmetric tensor, a single non-zero 
component of $u_{abc}$ brings out three-dimensional 
subspace by permutation of indices. 
Thus, $(10-d)$ dimensional part would naturally be decomposed 
into multiples of three-dimensional blocks. 
Furthermore, those blocks are subjected to the permutation 
symmetry under the interchange with each other. 
In this way, 
$\SO(d)\,(d=1 \cdots 7)$ symmetric ansatz has been investigated in
former works.   

Among those choices shown above, we examine $d=4$ and $d=7$ cases 
in particular. 
It is reported in Ref.~\cite{Kawai:2002jk} 
that $d=5$ and $6$ cases reduce to $\SO(7)$ ansatz, 
while $d=2$ and $3$ cases to $\SO(4)$ ansatz. $d=1$ had no solution. 

The preserved symmetry and the explicit forms of the full  
propagators for $d=4$ and $d=7$ cases are given as follows.

\noindent
{\bf SO(7) ansatz: } 
$\SO(7) \times \SO(3)$
\begin{equation}
	c_{ab} = {\rm diag}\bigl(\ 
		\text{7$c_1$'s},\ \text{3$c_2$'s}\ 
	\bigr) \,,
	\quad
	\sla{u} = u\,\Gamma^{8,9,10} \,,
\end{equation}

\noindent
{\bf SO(4) ansatz: } 
$\SO(4) \times \SO(3) \times \SO(3) \times Z_2$
\begin{equation}
	c_{ab} = {\rm diag}\bigl(\ 
		\text{4$c_1$'s},\ \text{6$c_2$'s}\ 
	\bigr) \,,
	\quad
	\sla{u} = \frac{u}{\sqrt{2}} 
		\bigl( \Gamma^{5,6,7} + \Gamma^{8,9,10} \bigr) \,,
\end{equation}
the $Z_2$ factor stands for the permutation symmetry between 
two $\SO(3)$ factors. 

For the ansatz presented above, we evaluate the free energy 
given by the sum of 1PI bubble diagrams. 
It is obtained as a series of the coupling constant $\lambda$ 
whose coefficients are functions of the parameters $m_1$, $m_2$, and $m$. 
Next, we calculate the second moment of eigenvalue distribution 
by differentiating the free energy with respect to $m_1$ and $m_2$. 
Finally, we apply the improved perturbation method to these 
perturbative series:
\begin{eqnarray}
	F^N (\lambda, m_1,m_2,u) \,
	&\longrightarrow& \,
	\widetilde{F}^N (\lambda, (m_0)_1,(m_0)_2,u_0) \nonumber \\
	&=&
	F^N (g\lambda, \{m_0\}+g(\{m\}-\{m_0\}))\Bigr|_{g^N, g\rightarrow 1,\{m\}=0},
\end{eqnarray}
$\{m\}$ and $\{m_0\}$ represent collectively the set of parameters 
$\{m_1, m_2, m\}$ 
and $\{(m_0)_1,(m_0)_2,m_0\}$, respectively. 
We put $m_1 = m_2 = m = 0$ so that it describes the original 
IIB matrix model. 
At this stage, we set $\lambda=1$. 
It is because 
the coupling constant $\lambda$ can be absorbed by 
the redefinition of the fields. 

We introduce the new variables $x_i$ as 
\begin{equation}
x_1 \equiv \frac{1}{(m_0)_1}, \quad
x_2 \equiv \frac{1}{(m_0)_2}, \quad
x_3 \equiv \frac{1}{m_0} \,,
\end{equation}
and use these parameters in the following section.

\subsection{Results in $\SO(4)$ ansatz}

In this subsection, 
we present the results in the case of $\SO(4)$ ansatz. 
We apply the improved perturbation method to the perturbative series 
for the free energy, the second moment of eigenvalue distribution 
for $X^a$ $(a=1 \cdots 4)$, which is denoted by $c_1$, 
and that for $X^a$ $(a=5 \cdots 10)$, which is denoted by $c_2$. 

As explained in Sec.~\ref{sec:optimized}, 
we first search for the extrema of these functions 
with respect to the artificial parameters $x_i$. 
Then, we find the accumulations of extrema and consider them 
as the candidates of plateau. 
We compute the weighted distribution of the functions in 
those regions. 
We use here the first derivatives (\ref{eq:1std}) and 
the second derivatives (\ref{eq:2ndd}) in particular. 
Finally, we identify the region as plateau in which 
two distributions overlap with each other. 
The graphs of distributions in individual regions are 
shown in Appendix~\ref{sec:detail}. 

\subsubsection{Free energy}

\begin{figure}
\includegraphics[scale=.8]{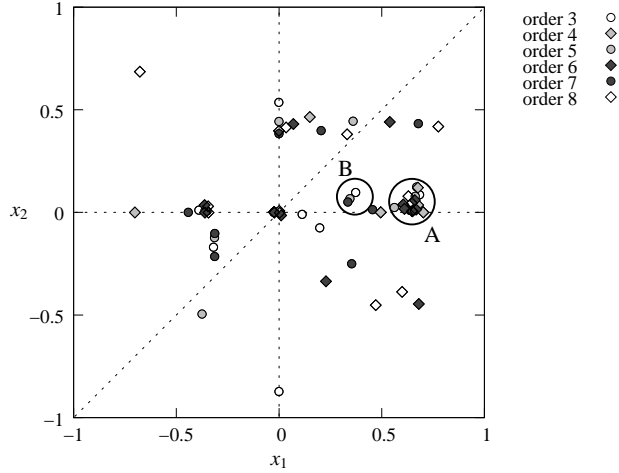}
\caption{%
Distribution of extrema of improved free energy for $\SO(4)$ ansatz 
plotted in $x_1$-$x_2$ plane.}
\label{fig:so4free01}
\end{figure}

Fig.~\ref{fig:so4free01} shows the distribution of extrema of 
the improved series of free energy. 
As is seen in Fig.~\ref{fig:so4free01}, 
there are several accumulations of extrema. 
We pick up some accumulations as the candidates of plateaux and compute
the distributions with weight functions $w_1$ and $w_2$. Besides most of  
accumulations show the hopeless distributions which spread randomly, 
two regions A and B which are depicted in Fig.~\ref{fig:so4free01} 
have a hopeful 
distribution. In the following, we refer to only hopeful candidates.

Two regions A and B are hopeful candidates of plateau, 
although the behavior of the function of the seventh order in region A  
is rather unstable. 
The similar situation also occurred in the previous work~\cite{Aoyama:2006rk} 
in which the positions of extrema show a different pattern 
for the seventh order of perturbation. 
In the region A we read the value of free energy to be -1 $\sim$ 2. 
And in the region B, we also read the value of free energy to be -1
$\sim$ 2.

From this accordance between region A and B about the value of free
energy, we may expect that two regions belong to one plateau. However 
from the examination of the other observables in the later subsections, 
we conclude that 
two regions are distinguished each others and there are two plateaux. 

\subsubsection{$c_1$}

\begin{figure}
\includegraphics[scale=.8]{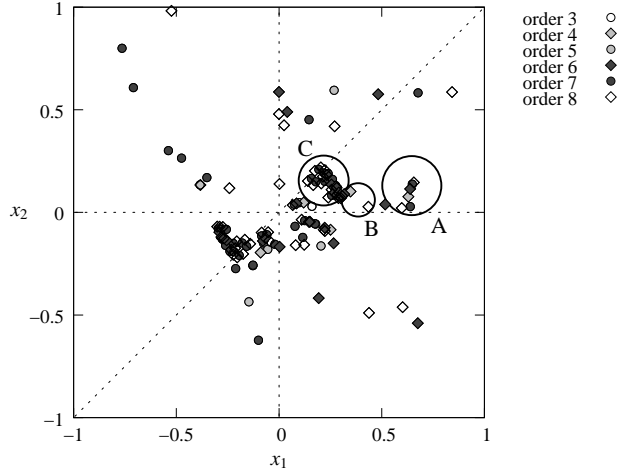}
\caption{%
Distribution of extrema of the improved $c_1$ for $\SO(4)$ ansatz.}
\label{fig:so4c101}
\end{figure}

Fig.~\ref{fig:so4c101} shows the distribution of extrema of 
the improved series of $c_1$. 
As mentioned in former section, 
although $c_1$ is the derivative function of free energy 
with respect to $m_1$, 
we have to apply the improved perturbation separately. 
It is because the behavior of the function becomes singular 
in the region of moduli space that corresponds to the massless case. 
It is seen by comparing 
Fig.~\ref{fig:so4free01} with 
Fig.~\ref{fig:so4c101}. 
Two distributions of extrema show the different pattern. 

Among accumulations, the regions A, B and C are 
promised candidates of plateau. 
The distribution in the region A shows a strange behavior. 
As the order of perturbation increases, the value at the peak
of distribution shifts to the right. 
When we approximate an infinitely large value by a finite series, 
such a behavior may often be observed. 
Therefore this property indicates that 
the exact value is quite large. 
Up to the eighth order of perturbation, 
we conclude that $c_1$ takes value of $40 \sim 60$. 
From the distribution in the region B, we read the value of $c_1$ 
to be $1.5\sim 2.1$.
Although regions A and B point the similar value of free energy, these 
regions shows clearly different distributions with respect to $c_1$. 
Therefore we conclude that these regions belong to different plateaux. 
On the other hand, from the region C, we read the value of $c_1$ 
to be $0.25\sim 0.26$. Thus we conclude that there are three plateaux.
In particular, it is interesting that the region C shows the 
same value as the region C of $c_2$ and $\SO(7)$ ansatz which will be
shown in later subsections. 
The significance of this accordance will be discussed later 
in Subsection~\ref{sec:so10}.

\subsubsection{$c_2$}

\begin{figure}
\includegraphics[scale=.8]{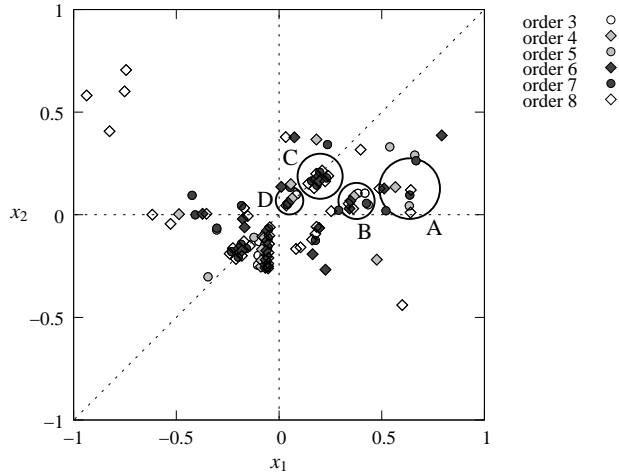}
\caption{%
Distribution of extrema of the improved $c_2$ for $\SO(4)$ ansatz.}
\label{fig:so4c201}
\end{figure}

Fig.~\ref{fig:so4c201} shows the distribution of extrema of 
the improved series of $c_2$. 
We consider the regions A, B, C and D as candidates of plateau. 
From the distribution in the region A, we read the value of $c_2$ 
to be $0.0\sim 0.2$. 
From the distribution in the region B, we read the value of $c_2$ 
to be $0.13 \sim 0.15$. 
From the distribution in the region C, we read the value of $c_2$ 
to be $0.25\sim 0.26$. 

In the similar fashion, region D indicates that $c_2$ takes the 
value $0.21 \sim 0.25$. 
From the similarity of the distributions in the region C and D, we
conclude that these two regions belong to one plateau. Thus we choose 
the region C as a representative of this plateau. 

\subsection{Results in $\SO(7)$ ansatz}

In this subsection, we show the results for the case of $\SO(7)$ ansatz. 
The method of analysis is the same as that in the $\SO(4)$ case. 
The graphs of distributions in individual regions are 
shown in Appendix~\ref{sec:detail}. 
 
\subsubsection{Free energy}

\begin{figure}
\includegraphics[scale=.8]{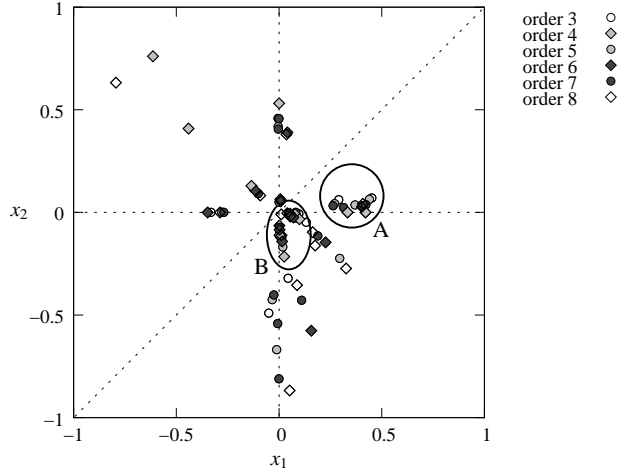}
\caption{%
Distribution of extrema of the improved free energy for $\SO(7)$ ansatz 
plotted in $x_1$-$x_2$ plane.}
\label{fig:so7free01}
\end{figure}

Fig.~\ref{fig:so7free01} shows the distribution of extrema of 
the improved series of free energy. 
Though it seems that the region B contains three distinct accumulations, 
we treat them as one region. 
It is because the distributions of three regions show the similar shape, 
then it is indicated that all three accumulations belong to one structure. 
Thus, there are two candidates of plateau. 
 
From the distribution in the region A, 
we read the value to be $5 \sim 7$.
From the distribution in the region B, 
we read the value of the free energy to be $-10 \sim 0$. 
The extrema in the region B are considered to be overshoots, 
because the weighted distribution with the second derivative 
deviates from that with the first derivative, 
and it is a characteristic feature of overshoot. 
Moreover, the behavior of the improved series 
of other observables around this region looks despairing. 
We conclude that it is indeed the overshoot.

\subsubsection{$c_1$}

\begin{figure}
\includegraphics[scale=.8]{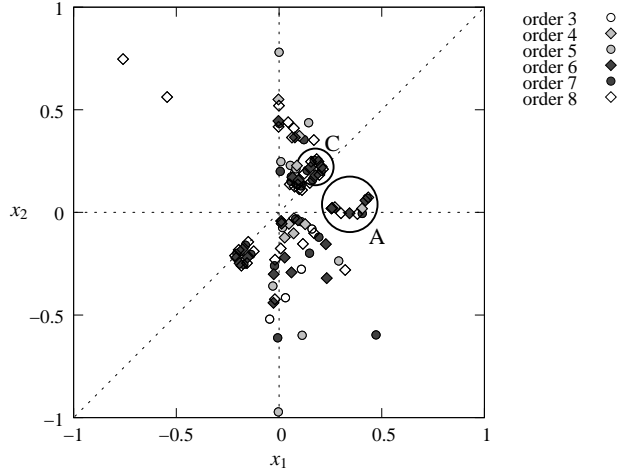}
\caption{%
Distribution of extrema of improved $c_1$ for $\SO(7)$ ansatz.}
\label{fig:so7c101}
\end{figure}

Fig.~\ref{fig:so7c101} shows the distribution of extrema of 
the improved series of $c_1$. 
Regions A and C are hopeful accumulations. 
From the distribution in the region A, 
we read the value of $c_1$ to be $0.7\sim 0.8$. 
From that in the region C, 
we read the value of $c_1$ to be $0.25\sim 0.26$. 

\subsubsection{$c_2$}

\begin{figure}
\includegraphics[scale=.8]{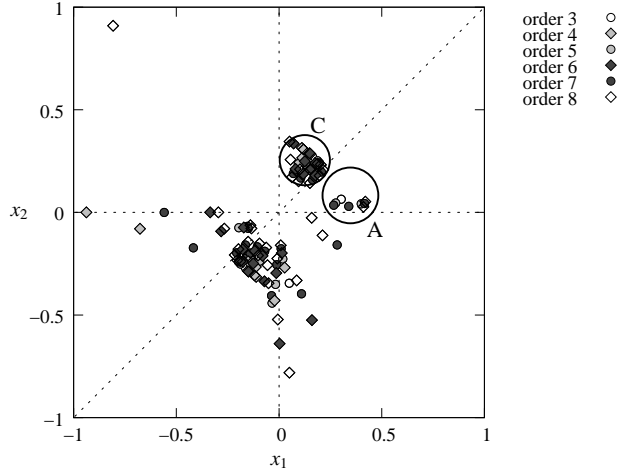}
\caption{
Distribution of extrema of improved $c_2$ for $\SO(7)$ ansatz.}
\label{fig:so7c201}
\end{figure}

Fig.~\ref{fig:so7c201} shows the distribution of extrema of 
the improved series of $c_2$. 
Regions A and C are hopeful regions. 
From the distribution in the region A, 
we read the value of $c_2$ to be $0.08\sim 0.12$. 
From that in the region C, 
we take the value of $c_2$ to be $0.25\sim 0.26$.

\subsection{$\SO(10)$-symmetric vacuum\label{sec:so10}}

From the investigation of the observables $c_1$ and $c_2$, 
we find regions in both $\SO(4)$ and $\SO(7)$ ansatz 
that have an interesting property. 
It is shown by region C 
in Figs.~\ref{fig:so4c101} and \ref{fig:so4c201} for $\SO(4)$ ansatz, 
and 
in Figs.~\ref{fig:so7c101} and \ref{fig:so7c201} for $\SO(7)$ ansatz 
as well. 
The approximate values of $c_1$ and $c_2$ estimated 
from the distributions 
are $0.25 \sim 0.26$, and they are almost the same. 
In addition, the region C is located at $x_3 \sim 0$, and thus 
it implies that the bi-fermionic variable $u_1$ vanishes. 
Therefore, we identify this region as the plateau 
that corresponds to the $\SO(10)$-symmetric vacuum. 

\begin{figure}
\subfigure[][]{%
	\includegraphics[scale=.85]{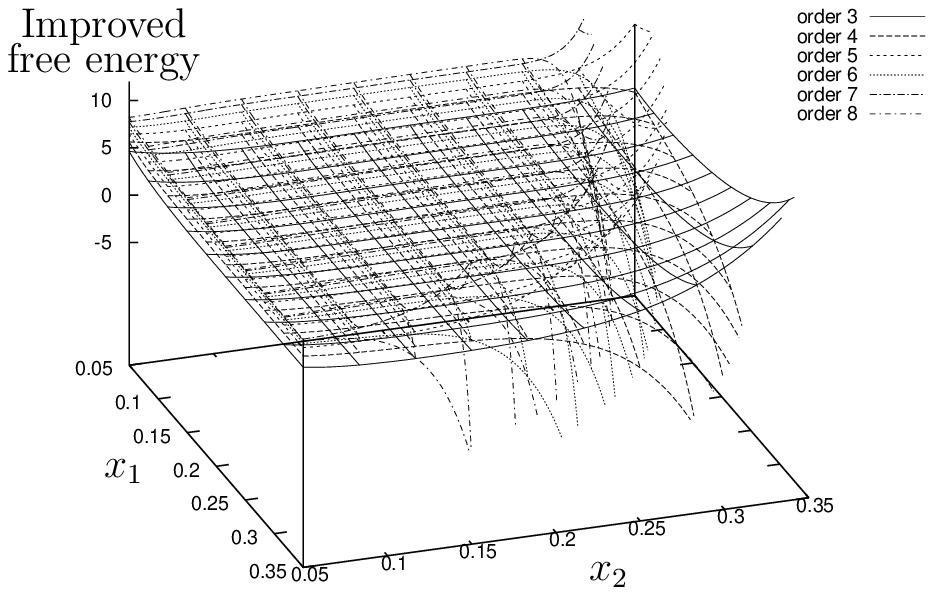}
	\label{fig:so10-1}
}
\subfigure[][]{%
	\includegraphics[scale=.75]{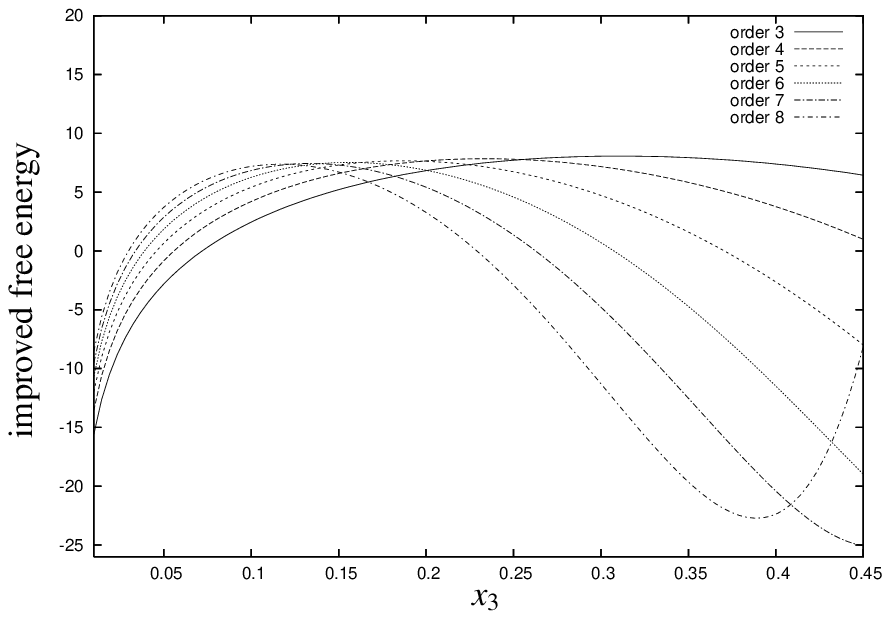}
	\label{fig:so10-2}
}
\caption{%
Dependence of the improved free energy with respect to $x_1$ and $x_2$ 
at $x_3=0.1$ \subref{fig:so10-1}, 
and $x_3$-dependence of the improved free energy 
at $x_1=x_2=0.15$ \subref{fig:so10-2} for $\SO(4)$ ansatz.}
\label{fig:so10}
\end{figure}

It may seem strange that there is no such accumulation of extrema 
found for the improved free energy around region C above. 
In fact, this region appears as ``smooth plateau'' discussed 
in Sec.~\ref{sec:plateau}. 
By plotting the dependence of the improved free energy 
(Fig.~\ref{fig:so10}), 
we can recognize that the improved free energy stays 
almost constant in region C, and thus the minimal sensitivity 
is indeed realized there. 
The approximate value of the free energy is obtained 
from the distribution shown in Appendix~\ref{sec:detail} 
as $6 \sim 7$. 

The above observations are obtained for 
both $\SO(4)$ ansatz and $\SO(7)$ ansatz, and 
the approximate values of the free energy, $c_1$, and $c_2$ 
are the same between those two ansatz. 
It is consistent with the speculation that 
these solutions obtained for two distinct ansatz are 
actually identical, in which the $\SO(10)$-symmetry is realized.

\subsection{Results in IIB matrix model} 

In the examinations of three observables, 
free energy, $c_1$ and $c_2$, 
for the $\SO(4)$ ansatz we find three sets of candidate of plateau.  

First plateau (we call this region as region A) indicates that 
the value of free energy is $-1 \sim 2$, 
that of $c_1$ reaches $50$ 
and that of $c_2$ is less than $0.2$ up to the eighth order of perturbation.  
This plateau corresponds to an anisotropic vacuum that preserves 
$\SO(4)$ subgroup. 
The ratio of extent of the eigenvalue distributions between 
the 4-dimensional part and the 6-dimensional part attains at $20$ 
at the stage of the eighth order of perturbation. 
The behavior of distributions as increase of a order
of perturbation indicates that the exact value of ratio is 
quite large. 
This solution has a property that the degree of 
anisotropy is large (probably infinite). 
Next plateau (we call this region as region B) indicates that 
the value of free energy is $-1 \sim 2$ 
and that of $c_1$ is $1.5 \sim 2.1$ 
and that of $c_2$ is $0.13 \sim 0.15$. 
This plateau also corresponds to an anisotropic vacuum that preserves 
$\SO(4)$ subgroup. 
The ratio of extent of the eigenvalue distributions between 
the 4-dimensional part and the 6-dimensional part is obtained 
by $3.1 \sim 4.0$. This solution has a property that the degree of
anisotropy is finite. 

The third candidate appears in a different manner. 
This is the plateau corresponding to a $\SO(10)$-symmetric vacuum 
which was discussed in the previous subsection.
For the observables $c_1$ and $c_2$, 
it indicates that both of $c_1$ and $c_2$ take 
the value $0.25 \sim 0.26$. 
For the free energy this plateau appears as ``smooth plateau''. 
By computing the distribution of free energy 
we conclude that the approximate value of free energy is $6 \sim 7$.
The graphs of distributions in this regions are 
shown in Appendix~\ref{sec:detail}. 

Similar argument can be applied to the case of $\SO(7)$ ansatz. 
There are two plateaux; one of which indicates that 
the value of free energy is $5 \sim 7$ 
and that of $c_1$ is $0.7 \sim 0.8$ 
and that of $c_2$ is $0.08 \sim 0.12$, 
and it corresponds to an anisotropic vacuum that 
preserves $\SO(7)$ subgroup. 
The ratio of the extent between the 7-dimensional part and 
the remaining 3-dimensional part is $2.4 \sim 3.1$. 
The other plateau corresponds to a $\SO(10)$-symmetric vacuum. 
This plateau indicates that 
the value of free energy is $6 \sim 7$ 
and that of $c_1$ and $c_2$ is $0.25 \sim 0.26$. 
Those values are 
the same as those obtained for the third plateau of $\SO(4)$ ansatz. 
It is consistent with the speculation that they are actually 
identical, and $\SO(10)$ symmetry is realized.

\section{Conclusion\label{sec:conclusion}}

We have investigated the properties of the improved 
perturbation method in much detail. 
This method is a sort of systematic optimization of a series 
expansion obtained by the perturbation theory, 
achieved by reorganizing the series with the artificially 
introduced parameters. 
In particular, we focus on the realization of the principle of 
minimal sensitivity, which provides a condition for the 
parameters that the improved series should be least dependent 
on them. 
A region in the parameter space that satisfies the above 
condition is called as ``plateau''. 
The exact value of the series would be reproduced on the 
plateau. 
Thus the efficient scheme for the identification of the plateau 
and evaluation of the approximate value becomes a major issue 
in the application of the improved perturbation method. 

We proposed a new scheme for evaluating the improved series 
which yields good approximate values. 
It relies on the distribution profile of the series that 
is weighted by the first and the second derivatives 
with respect to the artificial parameters. 
By incorporating those weightings, the distribution becomes 
more sensitive to such a behavior that the series stays 
almost constant in a certain region of the parameters, 
i.e. the plateau condition be satisfied. 
It is also noted that this scheme works effectively 
even when there are a number of parameters. 

We insist that the improved perturbation method should be 
applied independently for each individual observable of interest. 
It is because the prescription of improvement is quite nonlinear 
and the formation of plateau may occur in a different 
manner especially at low orders of perturbation. 
If they are supposed to coincide at high enough orders, 
the information of plateau of other observables may help to 
clarify the behavior of the improved series. 
Thus those information should be referred synthetically. 

We studied the IIB matrix model by the improved perturbation method 
in order to examine the non-perturbative solutions of the model. 
We concentrate on the possibility of the spontaneous breakdown 
of Lorentz symmetry in the present article. 
We applied our scheme to the analysis of the improved series 
up to eighth order of perturbation for 
$\SO(4)$ and $\SO(7)$ ansatz that have been obtained in former works. 
As a result, we found a new solution that corresponds to 
the ten-dimensional universe besides the four- and seven-dimensional 
solutions already found previously. 
A key to this detection resides in that we performed the analysis 
independently for individual observables and combined the information 
on the plateau of them. 

The existence of the plateau of the $\SO(10)$-symmetric vacuum 
is quite substantial because it provides an evidence 
for the validity of application of the improved perturbation 
method to the IIB matrix model. 
It ensures us that the investigation encloses the whole moduli 
space of the model including the $\SO(10)$ symmetry 
and that the improved series indeed describes the original 
$\SO(10)$-symmetric IIB matrix model. 
It is a non-trivial issue when the model exhibits phase transitions, 
as was discussed in the application to the Ising model \cite{Aoyama:2005nd}. 

Based on our new scheme for the improved perturbation method, 
we found that 
the value of free energy becomes large 
in the order of four-, seven- and ten-dimensional universe. 
Thus we conclude that 
the four-dimensional universe is more preferred than 
seven- and ten-dimensional universe. 

It is interesting that there are two distinct solutions 
that correspond to four-dimensional universe. 
One is the solution in which the degree of anisotropy is quite large, 
and the other is the solution in which the degree of anisotropy is 
finite. 
The estimated values of free energy are almost 
the same for those two solutions. 
It may suggest an interesting possibility 
that the inflation is caused by the phase transition 
from one SO(4) vacuum with finite degree of anisotropy 
to the other SO(4) vacuum that has infinitely large extent 
in four directions, i.e. our universe.

\begin{acknowledgments} 
The authors are grateful to H.~Kawai and T.~Matsuo 
for valuable discussions. 
Y.~S. is supported by 
the Special Postdoctoral Researchers Program at RIKEN. 
\end{acknowledgments} 

\appendix

\section{Details of analysis in IIB matrix model\label{sec:detail}}
In this appendix we show the graphs of distributions in individual
ansatz and these regions which are candidates of plateaux.

As we can see in Fig.\ref{fig:so4free01}-\ref{fig:so7c201}, for each
observables there are many accumulations of extrema. We must compute the 
distribution with weight function $w_1$, $w_2$ for each region. Most of 
regions show hopeless distributions. Then we select only the hopeful
regions here and depict the graphs of the distributions in these
regions. The locations of these regions in the artificial parameter
space are depicted in Fig.\ref{fig:so4free01}-\ref{fig:so7c201}. 

\subsection{SO(4) ansatz}
\subsubsection{Free energy}
Region A \\
\includegraphics[scale=0.95]{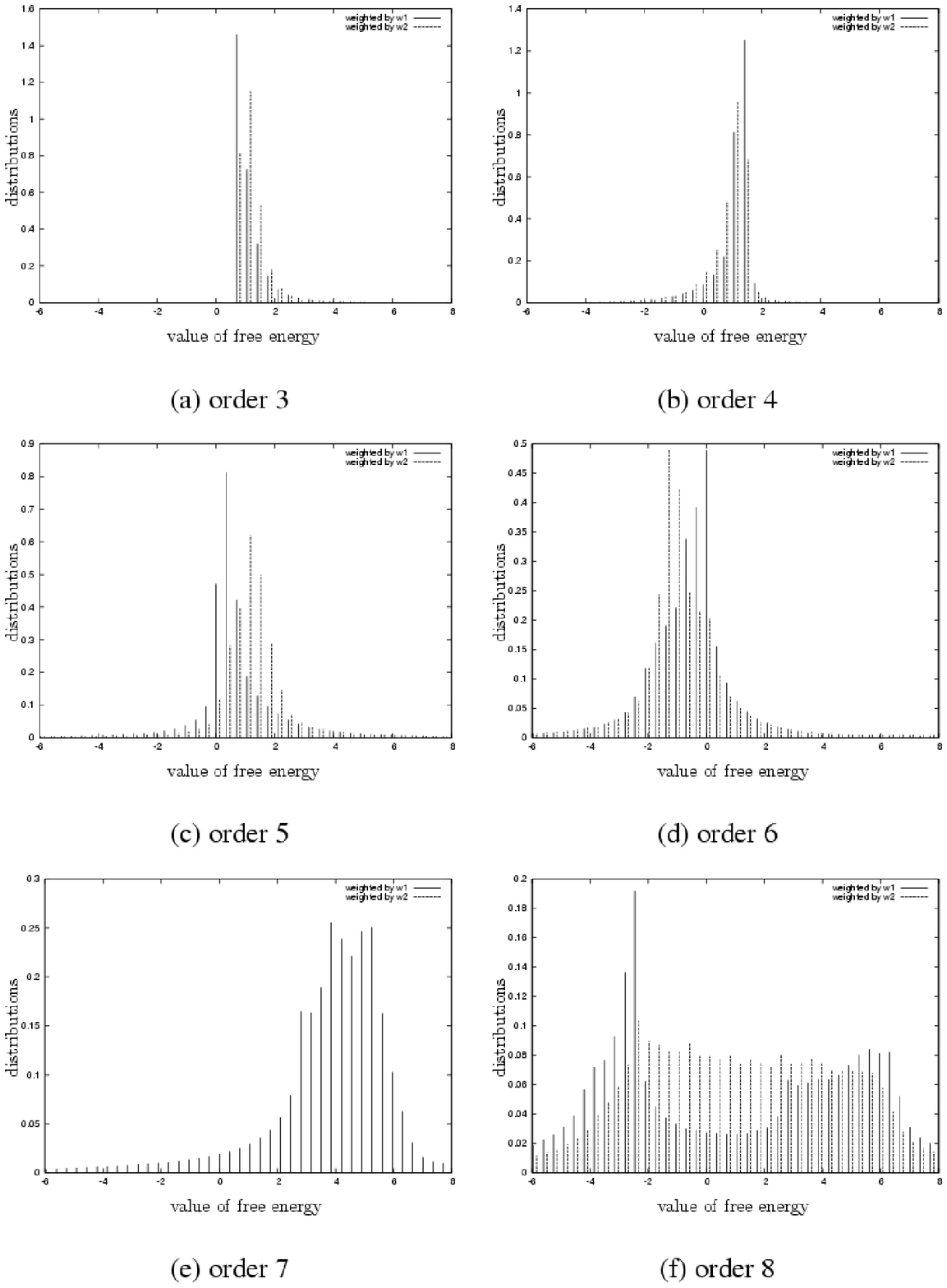}
\newpage
\subsubsection*{}
Region B \\
\includegraphics[scale=0.95]{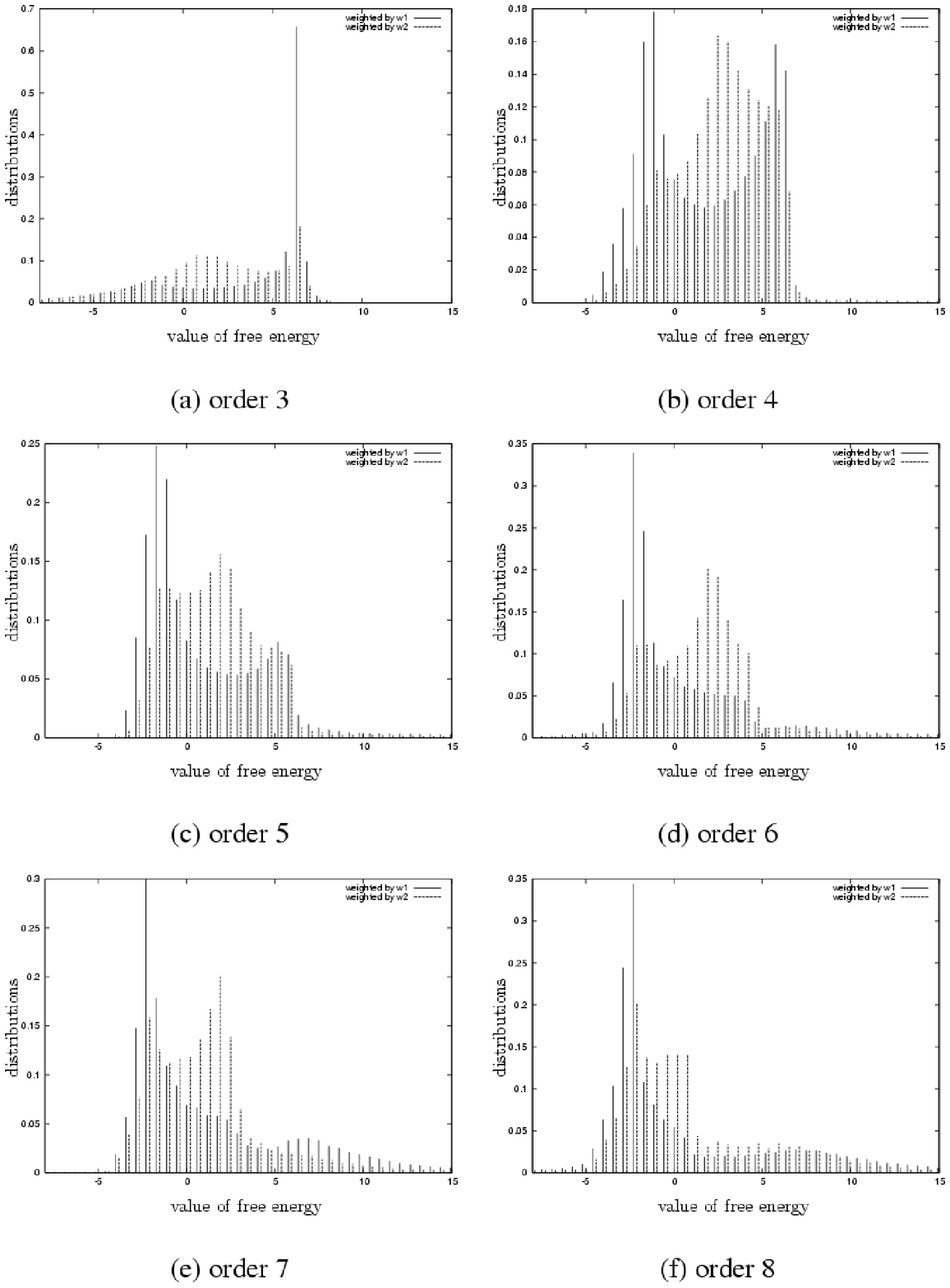}

\newpage

\subsubsection{$c_1$}
Region A \\
\includegraphics[scale=0.95]{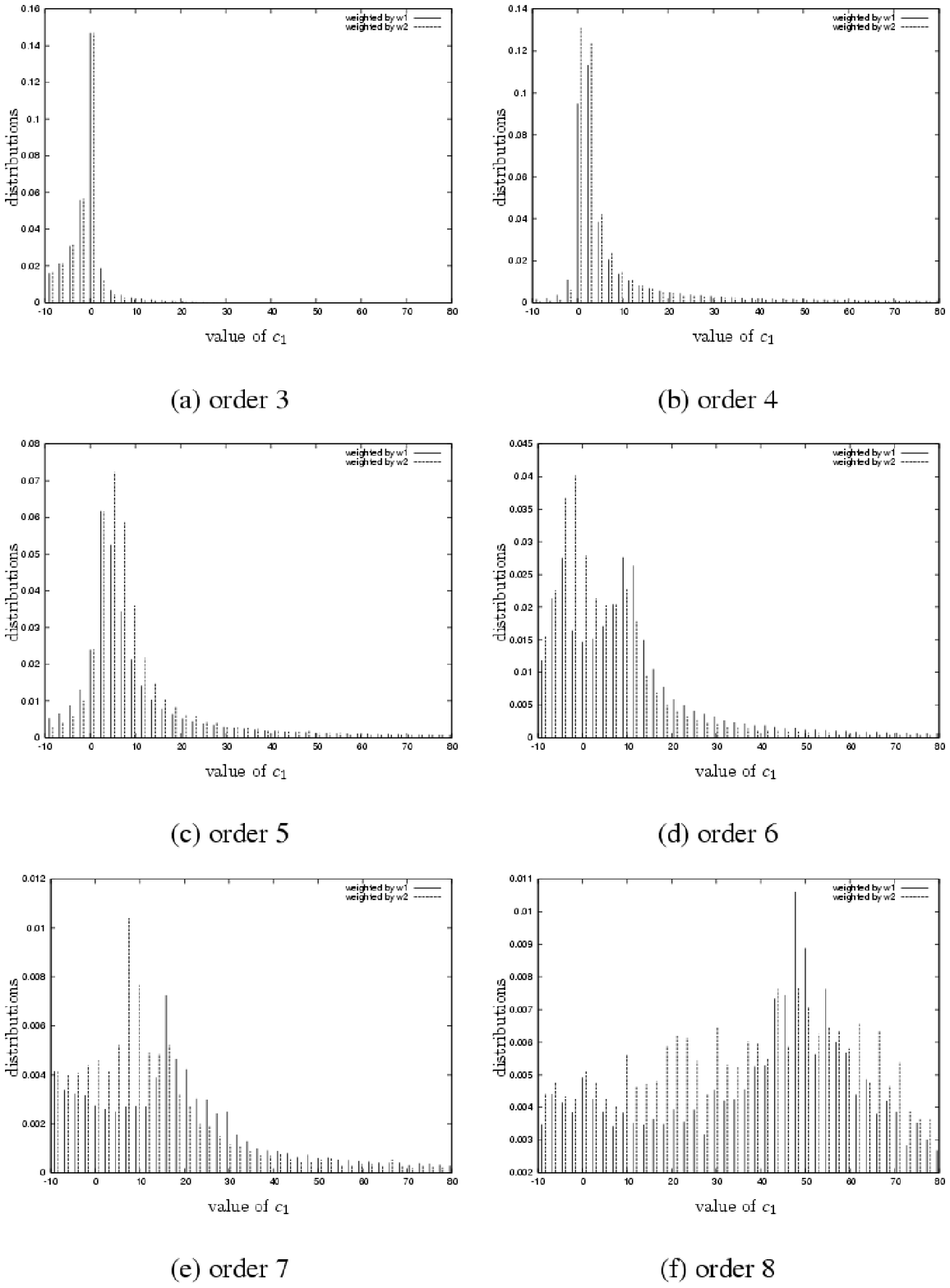}
\newpage
\subsubsection*{}
Region B \\
\includegraphics[scale=0.95]{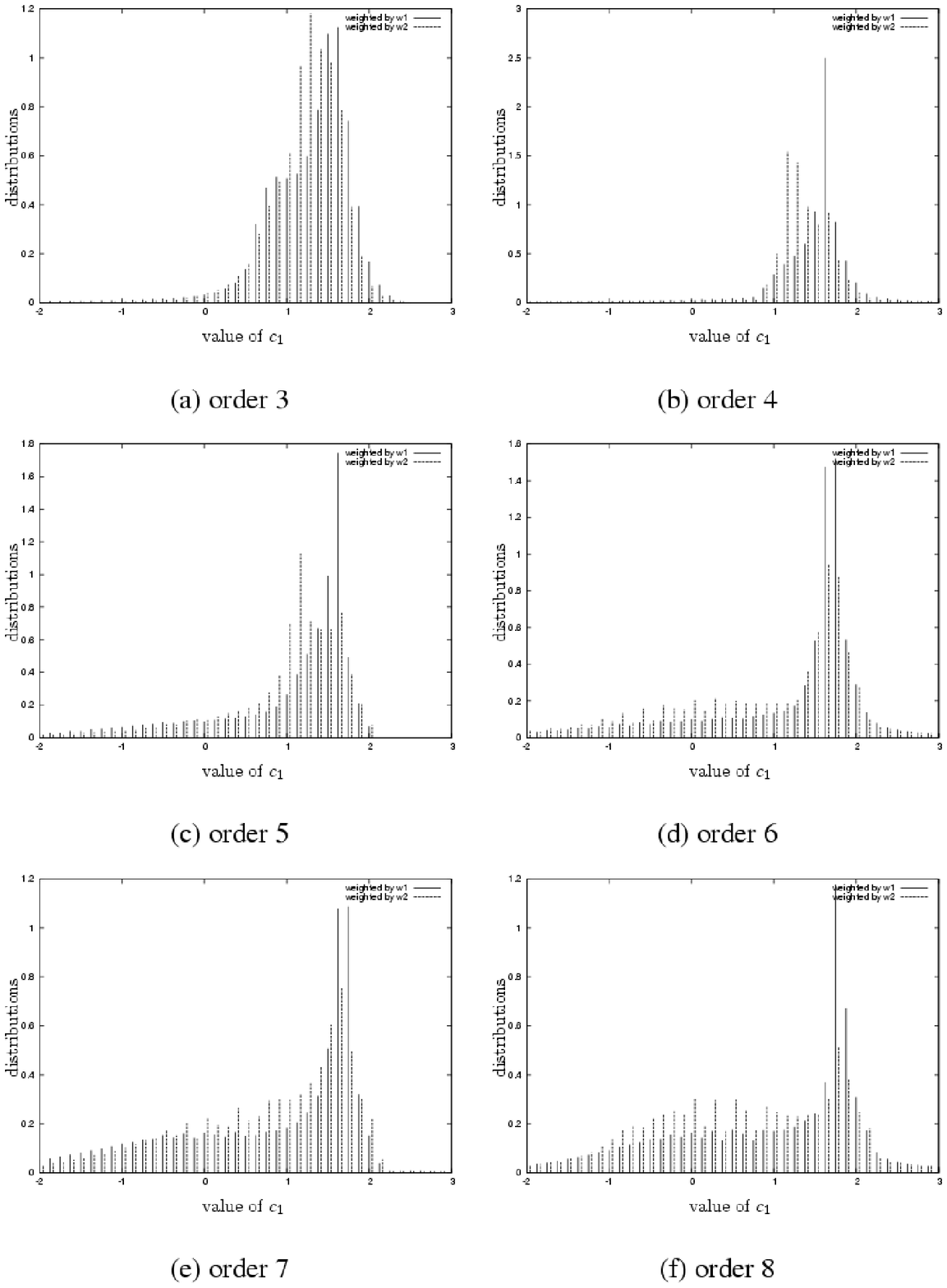}
\newpage
\subsubsection*{}
Region C \\
\includegraphics[scale=0.95]{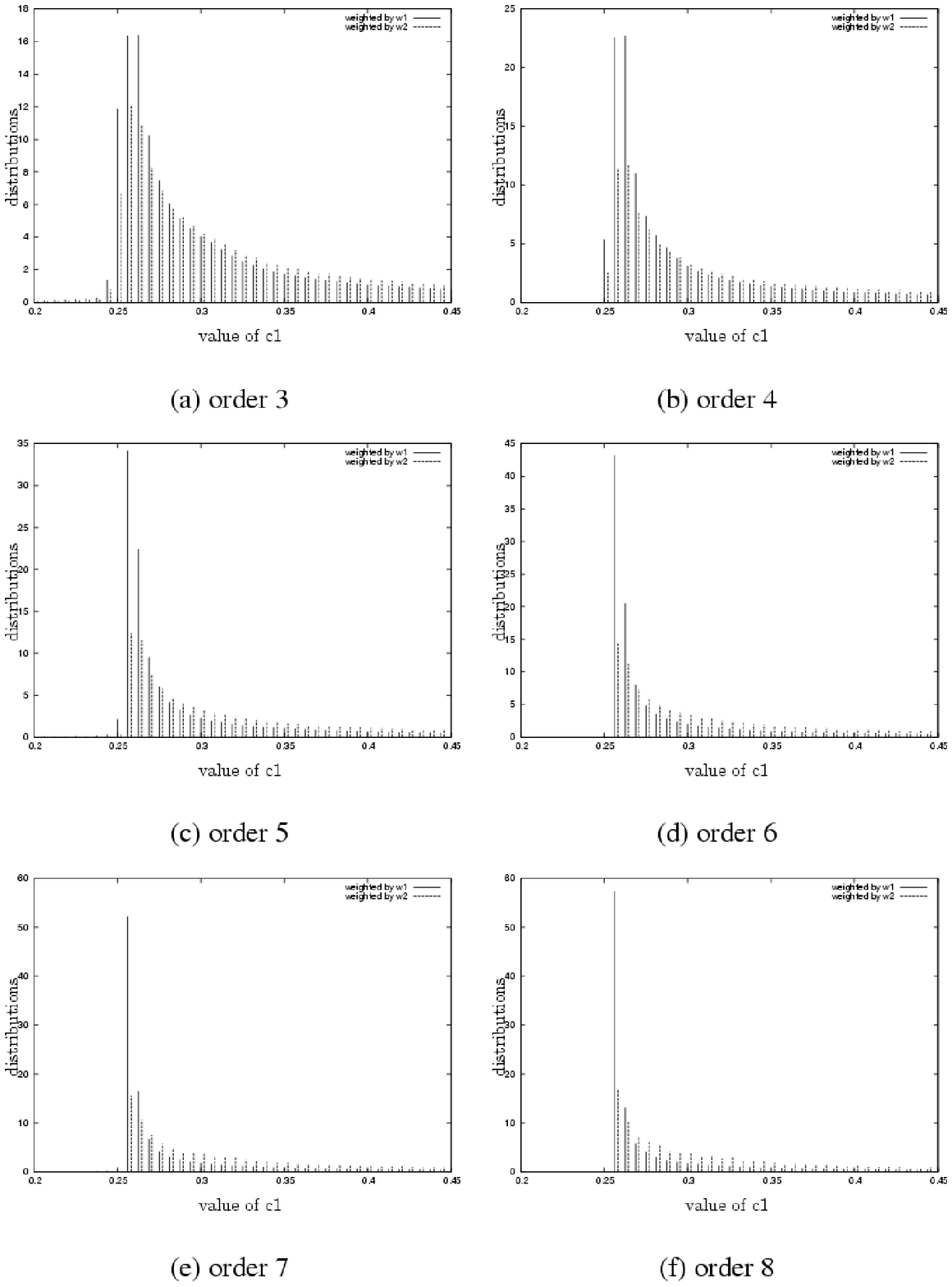}

\newpage

\subsubsection{$c_2$}
Region A \\
\includegraphics[scale=0.95]{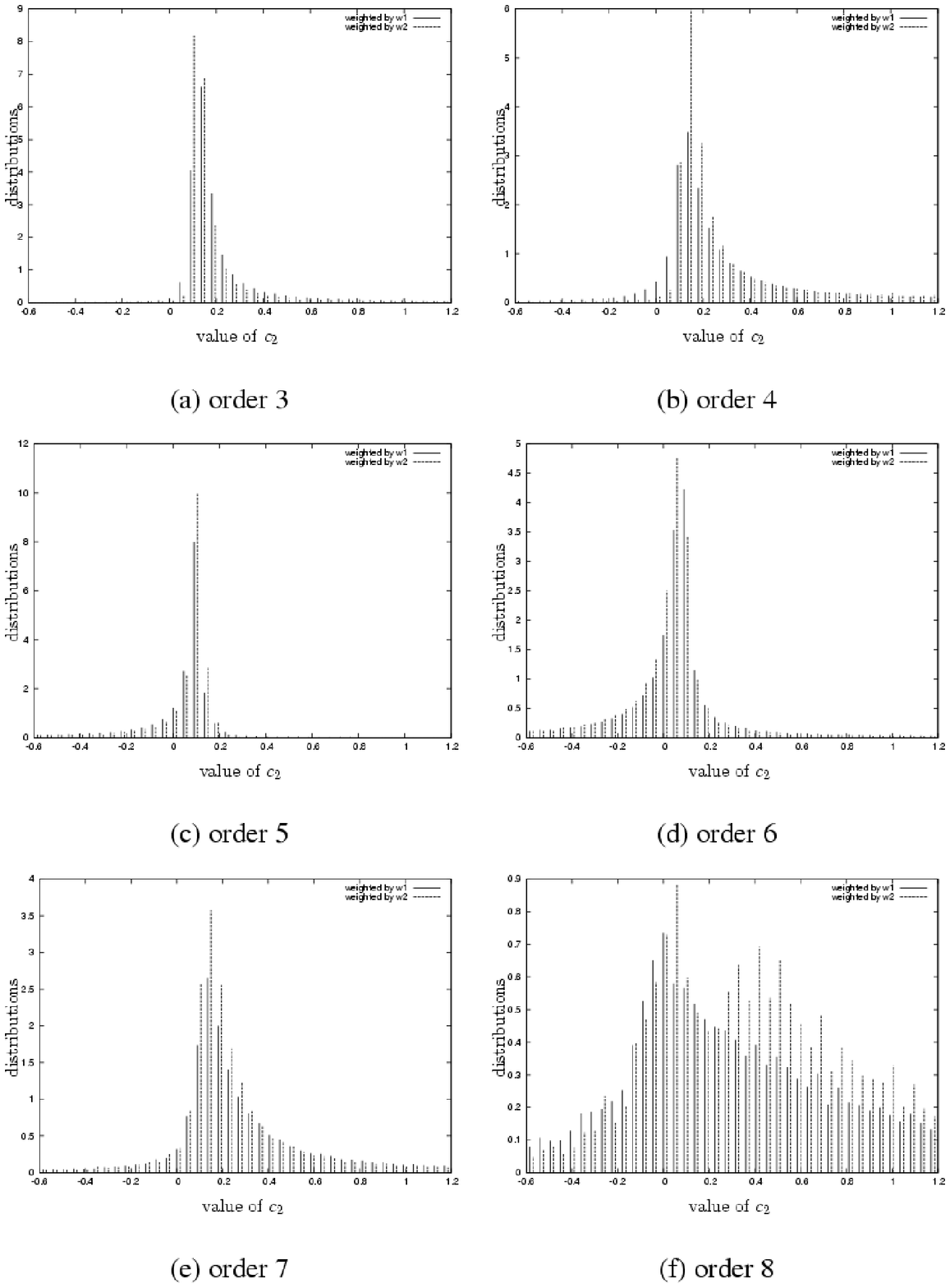}
\newpage
\subsubsection*{}
Region B \\
\includegraphics[scale=0.95]{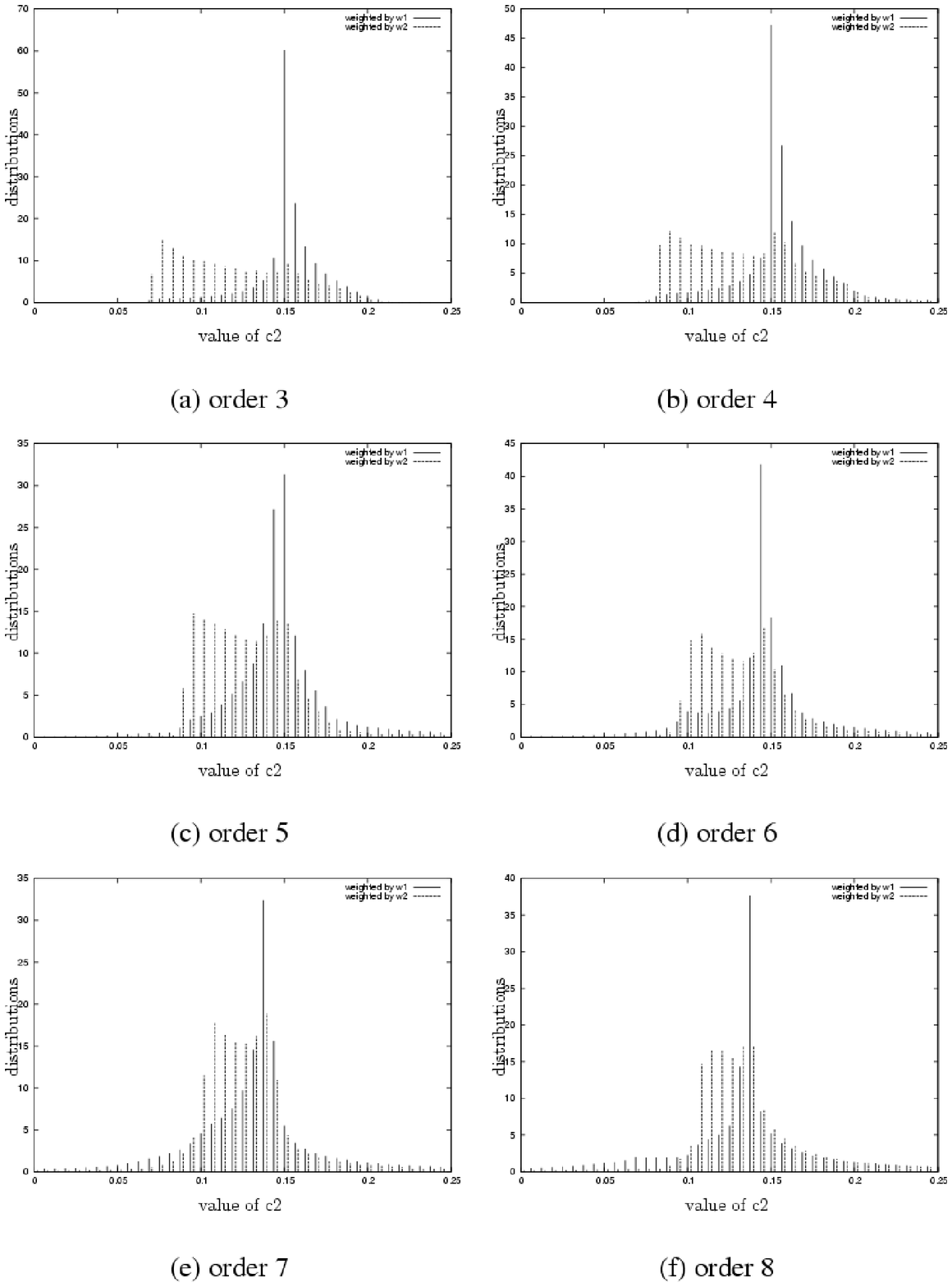}
\newpage
\subsubsection*{}
Region C \\
\includegraphics[scale=0.95]{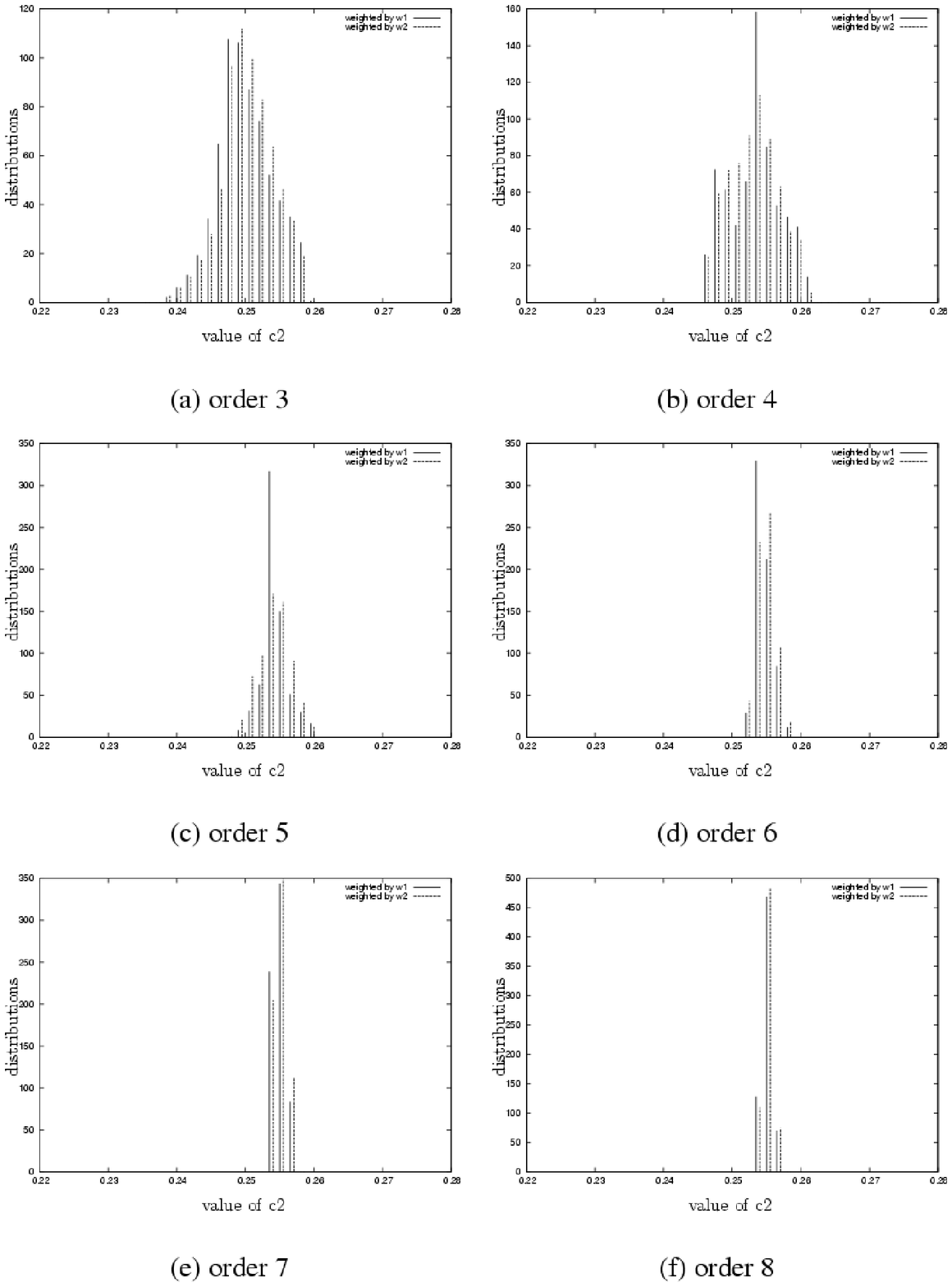}
\newpage
\subsubsection*{}
Region D \\
\includegraphics[scale=0.95]{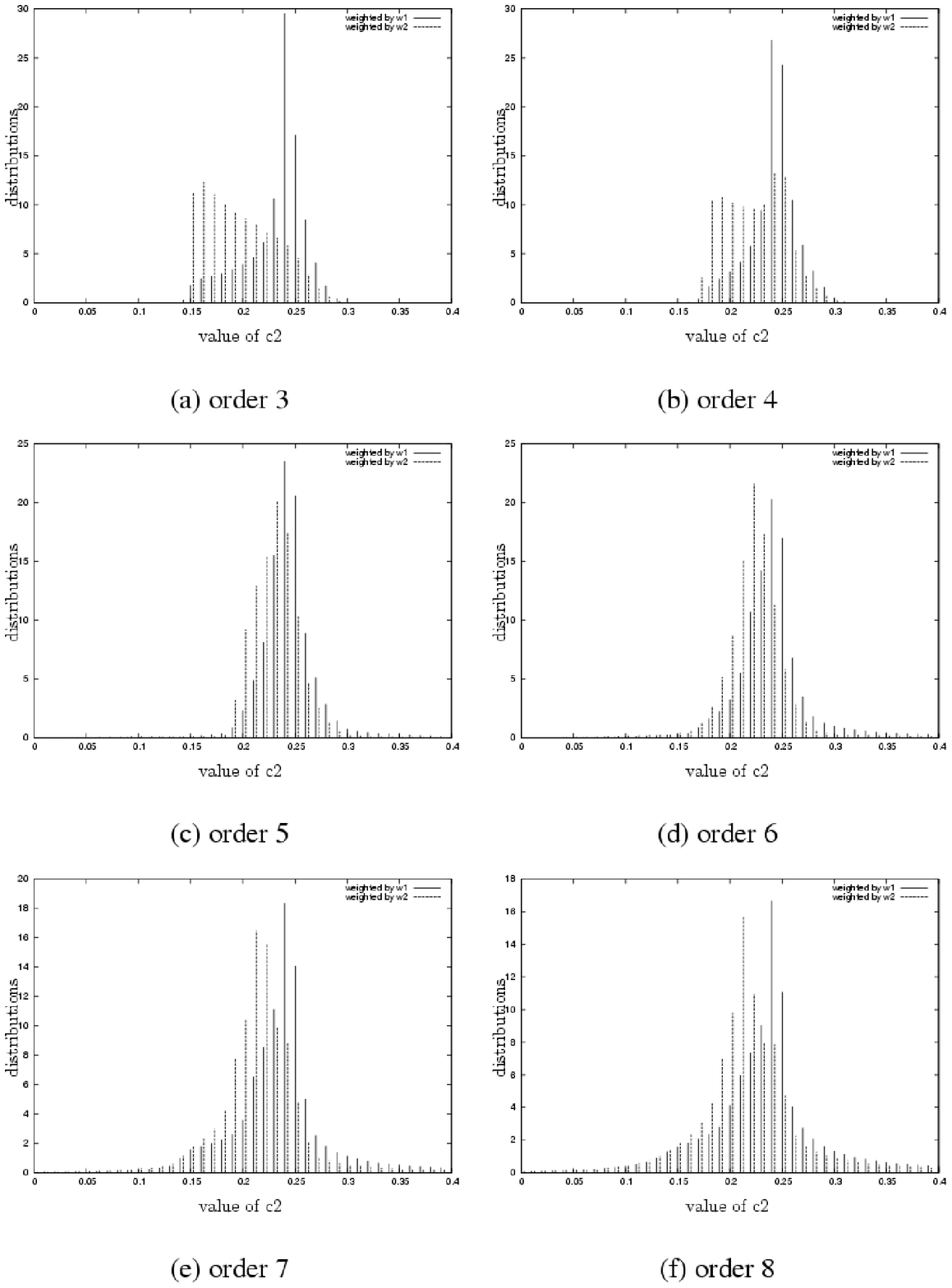}

\newpage

\subsection{SO(7) ansatz}
\subsubsection{Free energy}
Region A \\
\includegraphics[scale=0.95]{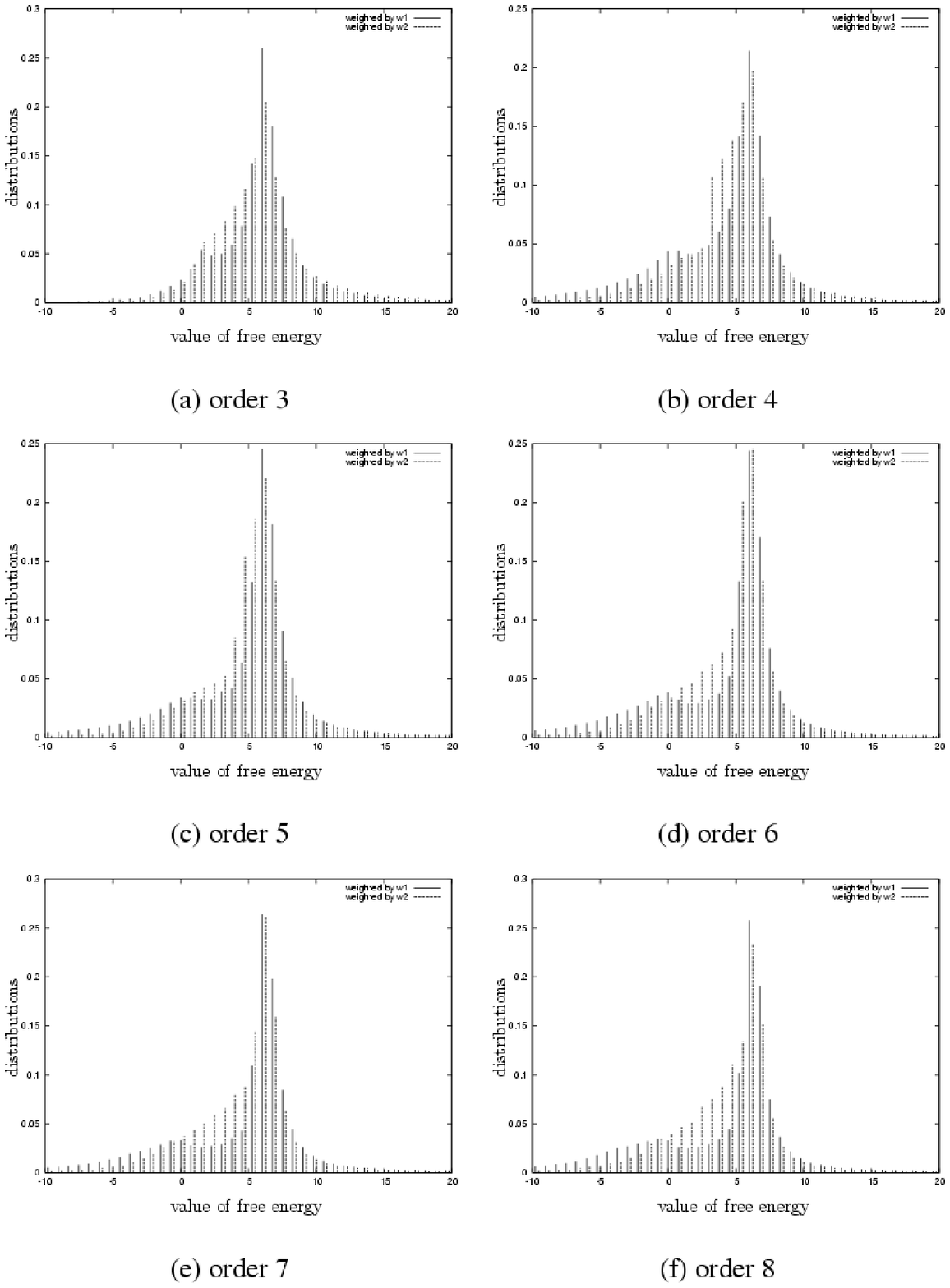}
\newpage
\subsubsection*{}
Region B \\
\includegraphics[scale=0.95]{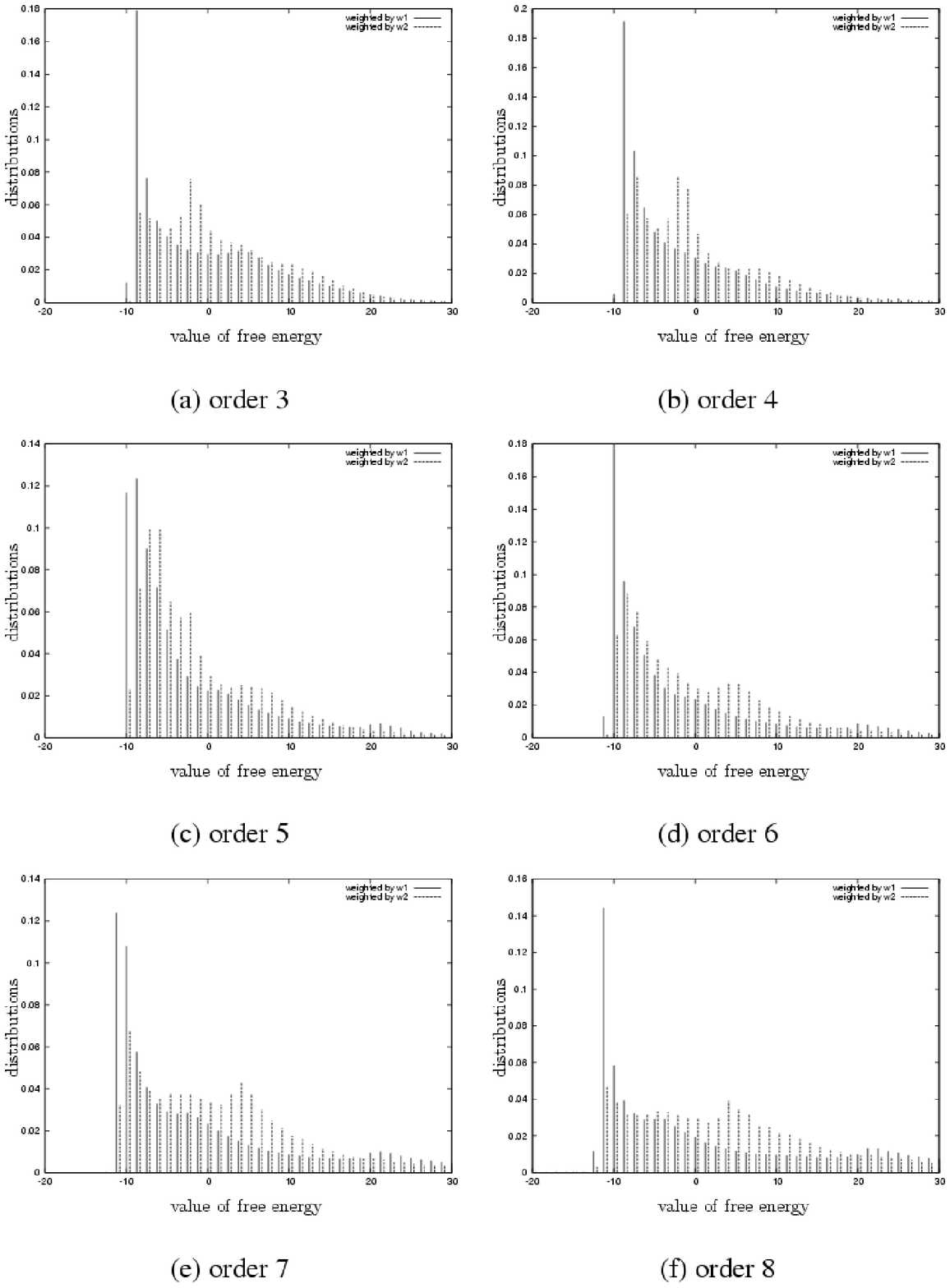}

\newpage

\subsubsection{$c_1$}
Region A \\
\includegraphics[scale=0.95]{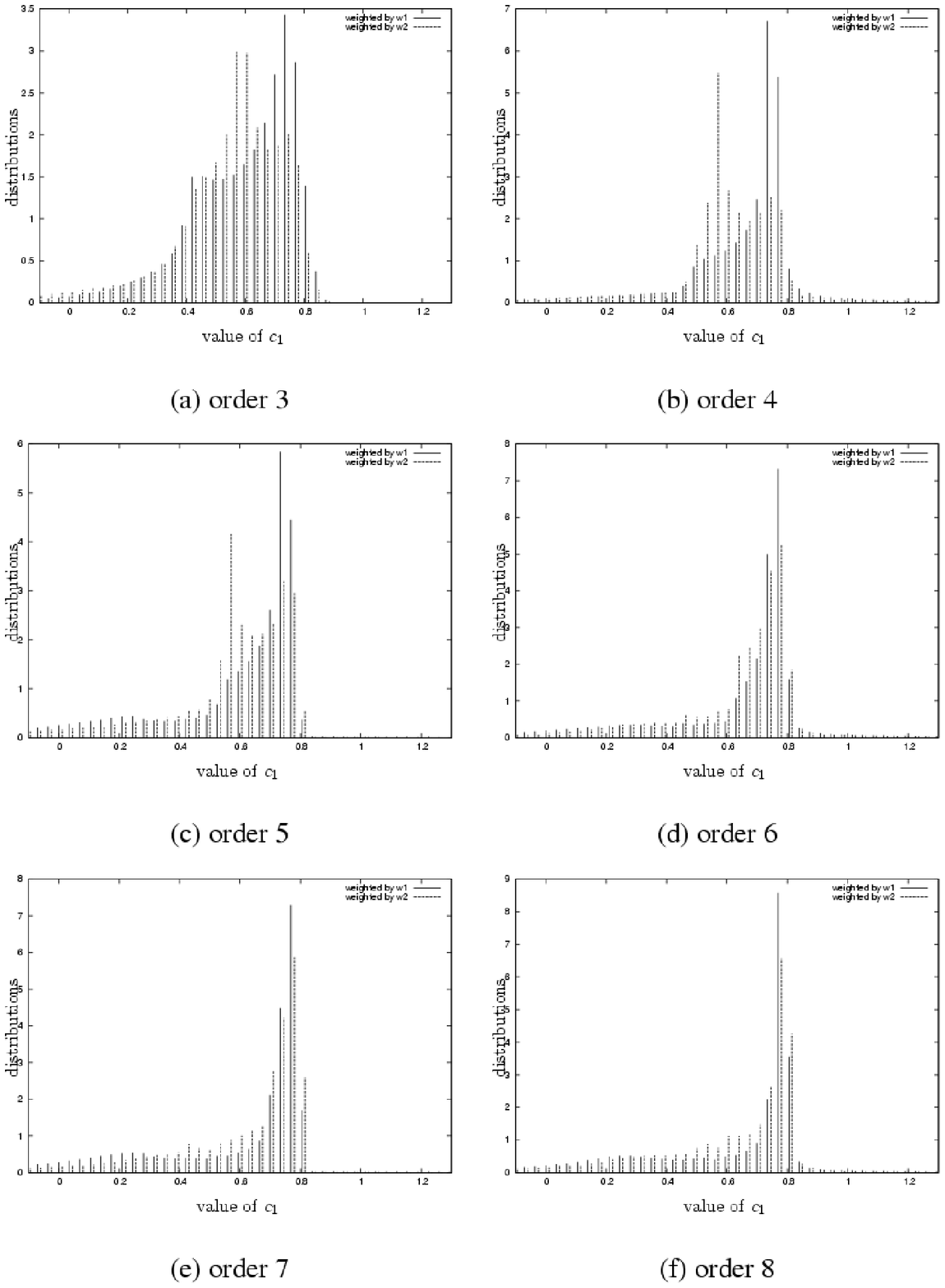}
\newpage
\subsubsection*{}
Region C \\
\includegraphics[scale=0.95]{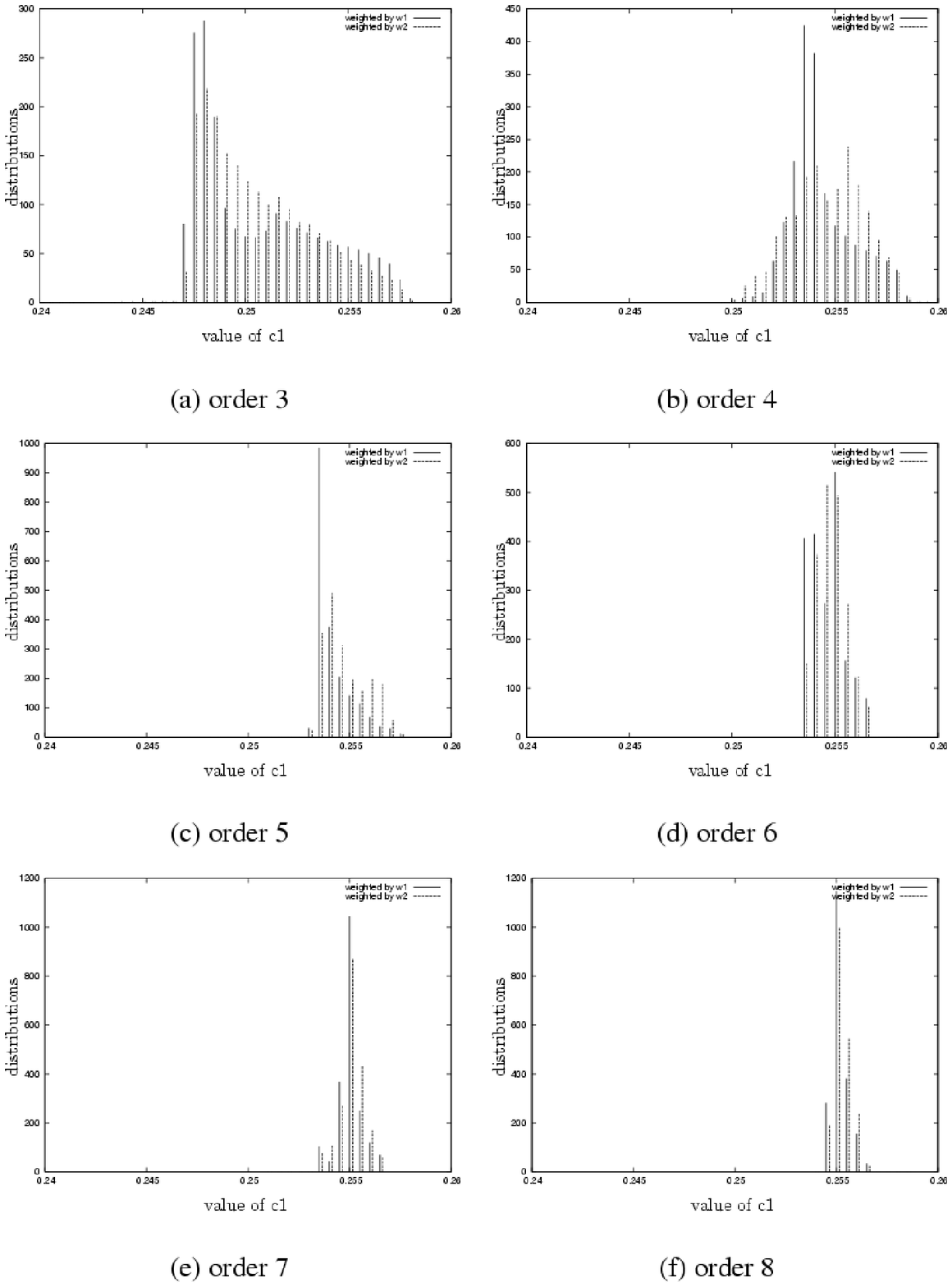}

\newpage

\subsubsection{$c_2$}
Region A \\
\includegraphics[scale=0.95]{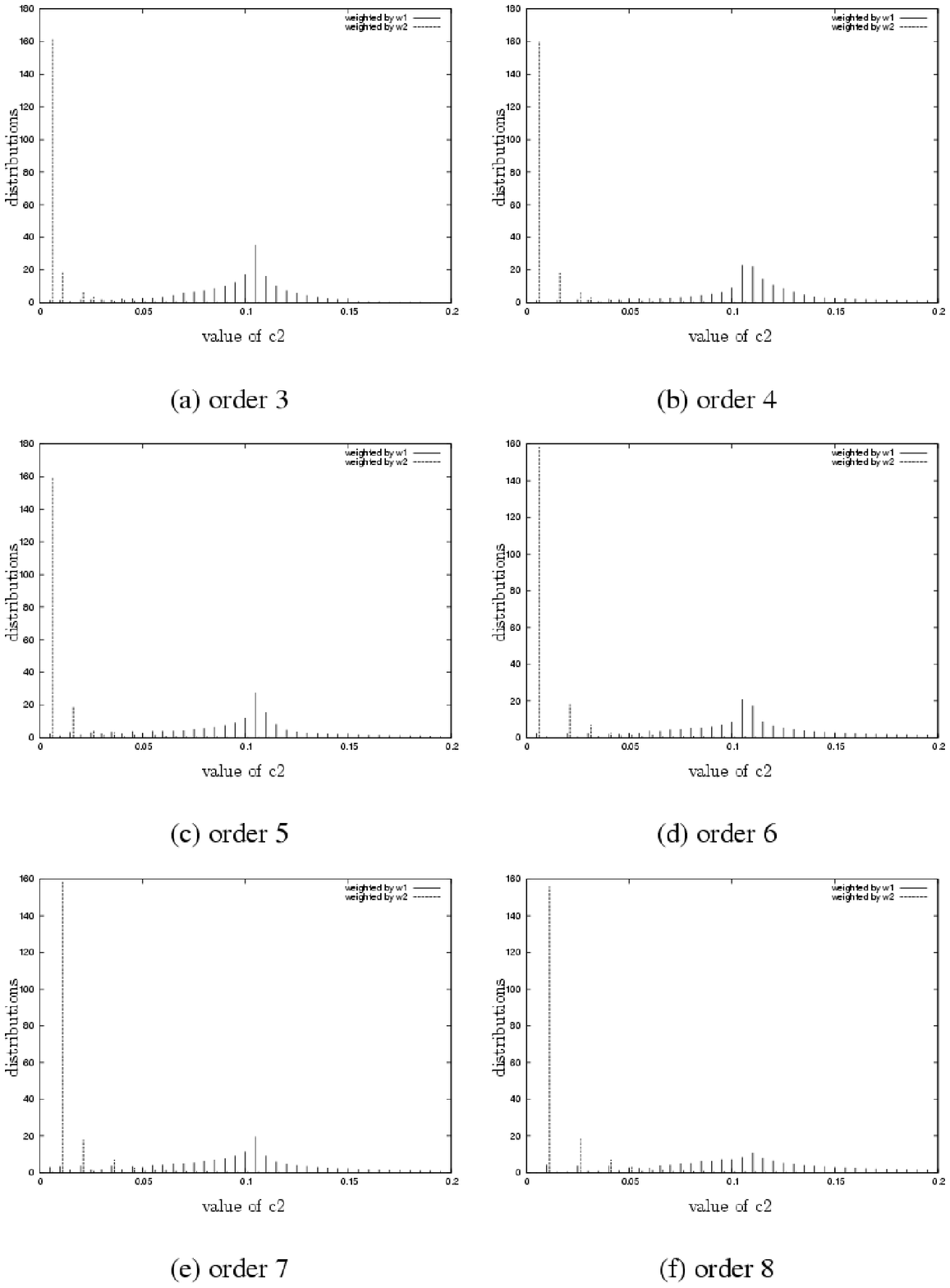}
\newpage
\subsubsection*{}
Region C \\
\includegraphics[scale=0.95]{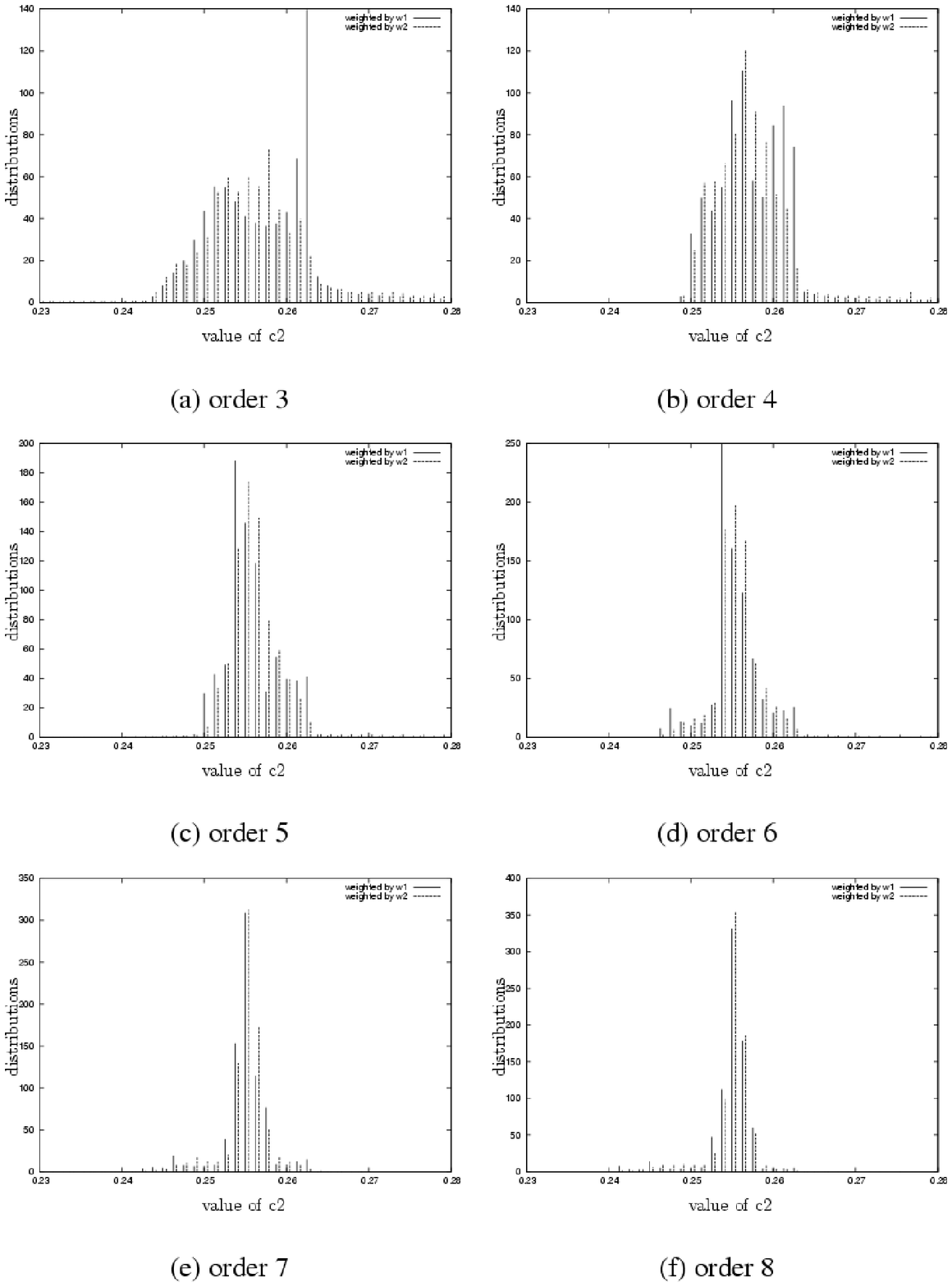}

\newpage

\subsection{SO(10)-symmetric vacua}
\subsubsection{Free energy}
Region C  \\
\includegraphics[scale=0.95]{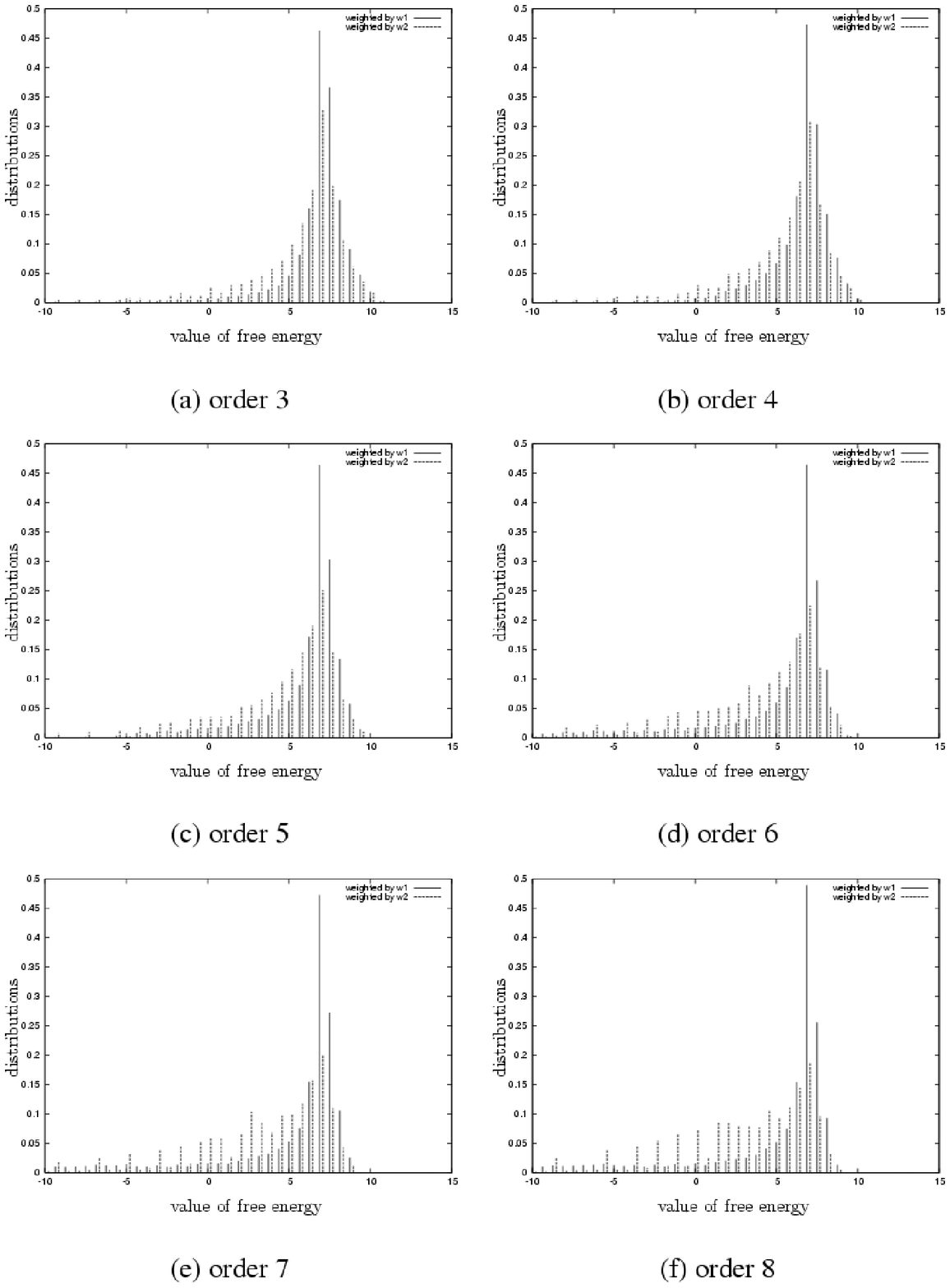}

\newpage




\end{document}